\documentclass[aps,prc,twocolumn,superscriptaddress,showpacs,nofootinbib]{revtex4-2}

\usepackage{graphicx}
\usepackage{longtable}
\usepackage{dcolumn}
\usepackage{bm}
\usepackage{ulem}
\usepackage{color}
\usepackage{epsfig}
\usepackage{natbib}

\newcommand{\al}{$\alpha$}
\newcommand{\g}{$\gamma$}
\newcommand{\raa}{($\alpha$,$\alpha$)}
\newcommand{\raX}{($\alpha$,$X$)}
\newcommand{\raXnull}{($\alpha$,$X_0$)}
\newcommand{\raXi}{($\alpha$,$X_1$)}
\newcommand{\raXLD}{($\alpha$,$X_{\rm{LD}}$)}

\newcommand{\rag}{($\alpha$,$\gamma$)}
\newcommand{\ran}{($\alpha$,n)}

\newcommand{\rann}{($\alpha$,2n)}

\newcommand{\rap}{($\alpha$,p)}

\newcommand{\rgn}{($\gamma$,n)}

\newcommand{\rga}{($\gamma$,$\alpha$)}

\newcommand{\stot}{$\sigma_{\rm{reac}}$}

\newcommand{\Nsv}{$N_A$$\left< \sigma v \right>$}

\newcommand{\rpro}{$r$-process}

\newcommand{\gpro}{$\gamma$-process}

\newcommand{\pnuc}{$p$-nucleus}
\newcommand{\sfact}{S-factor}

\newcommand{\smiv}{$^{144}$Sm}
\newcommand{\smvi}{$^{146}$Sm}
\newcommand{\gdvi}{$^{146}$Gd}
\newcommand{\gdvii}{$^{147}$Gd}
\newcommand{\gdviii}{$^{148}$Gd}
\newcommand{\euvii}{$^{147}$Eu}

\begin{document}

\title{
Cross section measurement of the $^{144}$Sm($\alpha$,n)$^{147}$Gd reaction for studying the alpha-nucleus optical potential at astrophysical energies
}
\author{Gy. Gy\"urky}%
\email{gyurky@atomki.hu}
\affiliation{Institute for Nuclear Research (ATOMKI), H-4001 Debrecen, Hungary}
\author{P. Mohr}%
\affiliation{Institute for Nuclear Research (ATOMKI), H-4001 Debrecen, Hungary}
\author{A. Angyal}%
\affiliation{Institute for Nuclear Research (ATOMKI), H-4001 Debrecen, Hungary}
\author{Z. Hal\'asz}%
\affiliation{Institute for Nuclear Research (ATOMKI), H-4001 Debrecen, Hungary}
\author{G.G. Kiss}%
\affiliation{Institute for Nuclear Research (ATOMKI), H-4001 Debrecen, Hungary}
\author{Zs. M\'atyus}%
\affiliation{Institute for Nuclear Research (ATOMKI), H-4001 Debrecen, Hungary}
\affiliation{University of Debrecen, H-4001 Debrecen, Hungary}
\author{T.N. Szegedi}%
\affiliation{Institute for Nuclear Research (ATOMKI), H-4001 Debrecen, Hungary}
\author{T. Sz\"ucs}
\affiliation{Institute for Nuclear Research (ATOMKI), H-4001 Debrecen, Hungary}
\author{Zs. F\"ul\"op}%
\affiliation{Institute for Nuclear Research (ATOMKI), H-4001 Debrecen, Hungary}

\date{\today}

\begin{abstract}
\begin{description}

\item[Background]
Nuclear reactions involving alpha particles play an important role in various
astrophysical processes such as the $\gamma$-process of heavy element
nucleosynthesis. The poorly known low-energy $\alpha$-nucleus optical
potential is a key parameter to estimate the rates of these reactions.

\item[Purpose]
The $\alpha$-nucleus optical potential can be tested by measuring the cross
section of $\alpha$-scattering as well as $\alpha$-induced reactions. Low
energy elastic $\alpha$-scattering on $^{144}$Sm has recently been
measured with high precision. The aim of the present work was to complement that work by measuring
the ($\alpha$,n) cross sections on $^{144}$Sm at low energies. The experimental data shall be
used to constrain the $\alpha$-nucleus optical model potential. From this
potential the $^{144}$Sm($\alpha$,$\gamma$)$^{148}$Gd reaction rate can be
derived with reduced uncertainties.

\item[Method]
The  $^{144}$Sm($\alpha$,n)$^{147}$Gd reaction was studied by bombarding Sm
targets with $\alpha$-beams provided by the cyclotron accelerator of
Atomki. The cross section was determined using the activation method. The
$\gamma$-radiation following the decay of the $^{147}$Gd reaction product was
measured with a HPGe detector. The experimental data are analyzed within the
statistical model.

\item[Results]
The cross section was measured in the $\alpha$-energy range between 13 and
20\,MeV in 1\,MeV steps, i.e., from close above the ($\alpha$,n)
threshold. The results were compared with statistical model calculations using
various approaches and parametrizations for the $\alpha$-nucleus optical
potential, and excellent agreement was obtained for two recent
potentials. However, these potentials cannot reproduce literature data for the
$^{144}$Sm($\alpha$,$\gamma$)$^{148}$Gd reaction with the same accuracy.

\item[Conclusions]
Constraints for the $\alpha$-nucleus potential were derived from an analysis
of the new $^{144}$Sm($\alpha$,n)$^{147}$Gd data and literature data for
$^{144}$Sm($\alpha$,$\gamma$)$^{148}$Gd. These constraints enable a
determination of the reaction rate of the
$^{144}$Sm($\alpha$,$\gamma$)$^{148}$Gd reaction with significantly reduced
uncertainties of less than a factor of two.

\end{description}
\end{abstract}

\pacs{
}

\maketitle

\section{Introduction}
\label{sec:intro}

The building-up of chemical elements by stars involve many different processes during the various stages of stellar evolution. The final episode of a massive star's life is the core-collapse supernova explosion, which is the site of several nucleosynthesis processes, for example the astrophysical $\gamma$-process \cite{Arnould2003,Rauscher2013}. The $\gamma$-process -- which may also take place in thermonuclear supernovae \cite{Travaglio2011} -- is thought to be the main process responsible for the synthesis of the so-called p-isotopes. These isotopes are those heavy, proton rich species - between $^{74}$Se and $^{196}$Hg - which are not produced by neutron capture reactions in the s- \cite{Kappeler2011} and r-processes \cite{Cowan2021}.

The $\gamma$-process proceeds through $\gamma$-induced reactions in the high temperature environment of a supernova. The main reactions are the neutron emitting ($\gamma$,n) reactions which drive the material towards the proton rich isotopes. Charged particle emitting ($\gamma$,p) and ($\gamma,\alpha$) reactions are, however, equally important if one wants to estimate the elemental and isotopic abundances resulting from a $\gamma$-process event.

Owing to the high number of reactions that participate in a $\gamma$-process network involving mostly unstable nuclei, experimental data for the reaction cross sections are very scarce and the astrophysical models are mainly based on theoretical reaction rates obtained from calculated cross sections. It has been found that in the case of reactions involving $\alpha$-particles, the cross sections are very sensitive on the choice of the $\alpha$-nucleus optical model potential (AOMP) and on its parameters. At low, astrophysically important energies (deep below the Coulomb-barrier) differences of up to two orders of magnitude are found between the various calculations. Moreover, the comparison with the very limited available experimental data shows that typically the calculations using global AOMPs are not able to reproduce the measured cross sections with the required precision. This introduces a large uncertainty into the astrophysical $\gamma$-process models.

The poor knowledge of AOMP necessitates its extensive experimental investigation. Traditionally, it has been studied with elastic $\alpha$-scattering experiments where the deviation from the Rutherford cross section carries information about the AOMP. However, in order to have large enough deviation, the experiments have to be carried out at relatively high energies, above the astrophysical energy range. Nevertheless, elastic scattering is still a useful tool, as it will be discussed later in the manuscript in relation to $^{144}$Sm.

The AOMP can also be studied by measuring the cross section of $\alpha$-induced nuclear reactions. Owing to their direct relevance for the $\gamma$-process, the radiative capture ($\alpha,\gamma$) reactions were studied in recent years on several isotopes (for a list of reactions see e.g. \cite{Szucs2019}). In the case of these reactions, however, the typically very low cross sections at astrophysical energies renders the measurements rather difficult. 

The problem related to the extremely low cross sections of the ($\alpha,\gamma$) reactions can be circumvented by measuring a reaction channel governed by the strong interaction. For example, ($\alpha$,n) reactions, if they are energetically possible, can also provide information about the AOMP and their cross sections are typically orders of magnitude larger than those of the radiative capture. Close above the neutron threshold the ($\alpha$,n) cross sections are sensitive mostly to the AOMP which can therefore be studied with these kind of reactions. For further details about the sensitivities on the various parameters see e.g. \cite{Rauscher2011}.

The p-isotope of Samarium, $^{144}$Sm, plays a special role in the history of $\gamma$-process studies. This was the first isotope on which ($\alpha,\gamma$) cross section measurement has been carried out in relation to the $\gamma$-process and the problem with the AOMP has been identified \cite{Somorjai1998}. A recent experiment has confirmed those results \cite{Scholz2020}. Elastic $\alpha$-scattering experiment on $^{144}$Sm has been carried out very recently at $\alpha$-energies of 16, 18 and 20\,MeV \cite{Kiss2022}. The AOMP models have been tested with these scattering data and it has been found that even at 16\,MeV the deviation from the Rutherford cross section is not large enough to draw conclusions with high confidence.

The threshold of the $^{144}$Sm($\alpha$,n)$^{147}$Gd reaction is at 12.6\,MeV
in the laboratory system. Above this energy, the
$^{144}$Sm($\alpha$,n)$^{147}$Gd can provide information about the AOMP of
$^{144}$Sm. There is only one data set for this reaction cross section in
literature by F.O.\,Denzler \textit{et al.} \cite{Denzler1995} but at the
lowest energies - near the threshold - there are only few data points which
bear high uncertainties both on the cross section value and on the energy
scale. Cumulative activation yields from another experiment by Archenti {\it
  et al.}\ \cite{Archenti1989} can be used to extract two further data points
for the \smiv \ran \gdvii\ reaction. But these data also have significant energy uncertainties of
1.5 MeV. All data from literature are thus not
suitable for a stringent test of AOMP models. The aim 
of the present work is therefore to measure the
$^{144}$Sm($\alpha$,n)$^{147}$Gd cross section with high precision from the
reaction threshold up to the energy range of the recent elastic scattering
experiment. The present results provide independent information about the
$^{144}$Sm + \al\ AOMP and complement the findings of the elastic scattering and
radiative capture experiments.  

\section{Experimental procedure}
\label{sec:experiment}

The reaction product of the $^{144}$Sm($\alpha$,n)$^{147}$Gd reaction is radioactive, decays with a half-life of 38.06\,$\pm$\,0.12~\,h to $^{147}$Eu \cite{NDS147}. The decay is followed by the emission of some high intensity $\gamma$-radiation. This allows the cross section measurement to be carried out by the $\gamma$-detection based activation technique \cite{Gyurky2019}, which was thus used in the present work. In the following sub-sections, the most important features of the experimental procedure are described. 

\subsection{Target preparation and characterization}
\label{subsec:target}

The p-isotopes of heavy elements have typically very low isotopic abundances, often below 1\,\%. $^{144}$Sm is an exception with its natural abundance as high as 3.07\,$\pm$\,0.07\,\% \cite{Rosman1998}. This fact, combined with the relatively large cross section of the ($\alpha$,n) reaction channel allowed the application of natural isotopic abundance targets. 

The targets were prepared by electron beam evaporation of natural isotopic composition samarium oxide (Sm$_2$O$_3$) onto 6\,$\mu$m thick Al foils. First information about the target thickness was obtained by weighing. The weight of the Al foils were measured to 1\,$\mu$g precision before and after the evaporation. Knowing the evaporated surface of 16\,mm in diameter, the target thickness could be calculated from the weight difference.

Since the weight does not give information about the target composition and the stoichiometry (i.e. the Sm:O ratio) which may change during the evaporation, the areal density of the Sm atoms - as the important quantity for the cross section determination - was measured with Rutherford Backscattering Spectrometry (RBS). For the RBS measurement a 2.0\,MeV $\alpha$-beam provided by the Tandetron accelerator of Atomki \cite{Rajta2018} was used. The scattered alpha particles were detected by a collimated ion-implanted Si detector mounted at a backward angle of 165 degrees with respect to the beam direction. The collected spectra were analyzed using SIMNRA \cite{SIMNRA}, a widely used computer code for simulating and evaluating - among others - RBS spectra. The thicknesses from the RBS measurement were used for the cross section analysis which were in agreement with the weighing data within 5\,\% when the Sm:O ratios obtained from the RBS analysis were used for the weight measurements. 

Altogether five targets were prepared and the Sm thicknesses were measured to be in the range of (0.5\,--\,1.0)$\cdot$10$^{18}$ atoms/cm$^2$ with a precision of 5\,\%. Each target was used one or two times in the subsequent cross section measurement campaign. Thicker targets were used for lower energies where the cross sections are lower. 

\subsection{Irradiations}

For the $^{144}$Sm($\alpha$,n)$^{147}$Gd cross section measurements, the targets were irradiated by $\alpha$-beams provided by the K20 cyclotron accelerator of Atomki \cite{Biri2021}. The irradiation chamber was the same as in our previous works, see e.g. \cite{Korkulu2018}. The whole chamber served as a Faraday cup, allowing the determination of the number of projectiles by charge integration. The He$^{++}$ beam intensity was typically 1\,e$\mu$A. 

The length of the irradiation varied between 2 and 24 hours, longer irradiation was used at lower energies where the cross section is smaller. During the irradiation the beam current was recorded in multichannel scaling mode with one minute resolution so that the fluctuations in the beam intensity could be taken into account in the activation analysis \cite{Gyurky2019}.

The cross section measurements were carried out in the energy range from 13\,MeV (close above the threshold) up to 20\,MeV with 1\,MeV steps. The highest energy coincides with the highest energy of the recent elastic scattering experiment on $^{144}$Sm \cite{Kiss2022}. 

With one exception, a single target was irradiated in a single activation run. In the case of the E$_\alpha$\,=\,20\,MeV irradiation, however, two targets were placed in the chamber, one behind the other, separated by a 10\,$\mu$m thick Al energy degrader foil, similar to the procedure of a recent experiment described in \cite{Gyurky2021}. The thickness of the degrader foil was measured with $\alpha$ energy loss to a precision of 5\,\% which was used for the effective energy determination at the second target (see section \ref{sec:results}). 

\subsection{Detection of the decay radiation}

The decay of the $^{147}$Gd reaction product is followed by the emission of many different energy $\gamma$-radiations from numerous cascade transitions. For the cross section determination, we have used the four most intense radiations with energies 229.2\,keV (57.7\,\%\,$\pm$\,3.7\,\%), 370.0\,keV (15.7\,\%\,$\pm$\,0.9\,\%), 396.0\,keV (31.4\,\%\,$\pm$\,1.9\,\%) and 929.0\,keV (18.4\,\%\,$\pm$\,1.1\,\%). In parentheses the relative intensities are given which were taken from the latest nuclear data compilation \cite{NDS147}.

After the irradiation the targets were removed from the chamber and transported to the off-line counting setup. This setup consisted of a LEPS detector (Low Energy Photon Spectrometer, a thin crystal high-purity germanium
detector optimal for the detection of low-energy $\gamma$-radiation) and a complete 4$\pi$ lead shielding with copper and cadmium liners \cite{Szucs2016}. 

In order to measure low activities, the targets were placed in close geometry in front of the detector, at a distance of 1\,cm from the crystal. In such a close geometry, the true coincidence summing effect makes it difficult to determine precisely the detection efficiency and the activity of the produced $^{147}$Gd source. Therefore, the so-called two-distance technique was used \cite{Gyurky2019}. The absolute efficiency of the detector was measured with calibrated radioactive sources at far geometry (in this case at 10\,cm from the detector) where the summing effect is negligible. Then the decay of a high activity $^{147}$Gd source was measured both in close and far geometries. Based on the two measurements, an efficiency conversion factor between the two geometries was obtained for all studied transitions, which was then used for the low activity samples that could be measured only in close geometry. 

\begin{figure}
\includegraphics[width=\columnwidth]{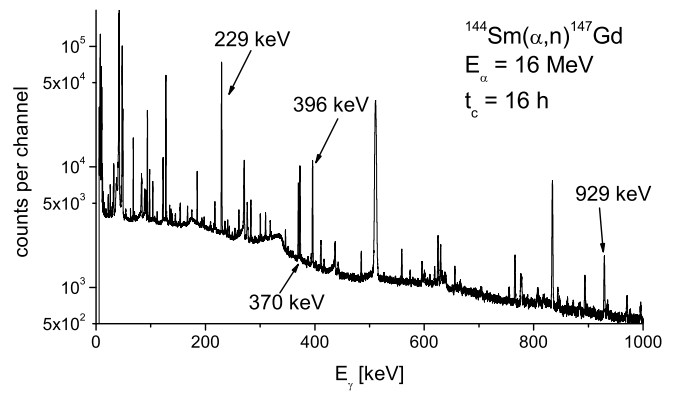}
\caption{\label{fig:spectrum} Typical $\gamma$-spectrum measured for 16\,hours after irradiating the target with a 16\,MeV $\alpha$-beam. The peaks corresponding to the transitions used for the cross section determination are marked. The channel width is about 0.1\,keV.}
\end{figure}

The length of the $\gamma$-counting was between 12 and 140 hours and the spectra were stored hourly in order to follow the decay of $^{147}$Gd. As an example, Fig. \ref{fig:spectrum} shows a $\gamma$-spectrum which was measured on a target irradiated with an $\alpha$-beam of 16\,MeV. The peaks corresponding to the transitions used for the cross section determination are marked.

\subsection{Results}
\label{sec:results}

Table\,\ref{tab:results} shows the measured cross section values. The first column contains the primary $\alpha$-beam energies provided by the cyclotron, while in the second column the effective interaction energies and their uncertainties are shown in the center-of-mass system. The effective energy calculation is based on the energy loss of the beam in the target layer. Since this energy loss is relatively small (a few tens of keV) and the cross section does not change much in such an energy interval, the effective energy was calculated for half of the target thickness. The uncertainty of the effective energy is made up by the uncertainty of the primary beam energy (0.3\,\% from the energy calibration of the cyclotron), the target thickness and the stopping power uncertainty. In the case of the measurement carried out with a degrader foil (marked with asterisk in the table), the energy uncertainty also includes the contribution of the degrader foil thickness.

The cross section results listed in the last column were calculated as the average of the values obtained from the four studied transitions, which were always in good agreement with each other. The uncertainty of the cross section is calculated as the quadratic sum of the following sources: target thickness (5\,\%), detection efficiency including the far-close geometry conversion factor (5\,\%), beam current integration (3\,\%), relative intensity of the $^{147}$Gd decay radiation (5\,\%), $^{144}$Sm natural abundance (2.3\,\%) and counting statistics ($<$5\,\%). 

\begin{table}
\caption{\label{tab:results} Measured cross section of the $^{144}$Sm($\alpha$,n)$^{147}$Gd reaction.}
\begin{tabular*}{\columnwidth}{@{\hskip\tabcolsep\extracolsep\fill}lr@{\hspace{-6mm}}c@{\hspace{-6mm}}lr@{\hspace{-6mm}}c@{\hspace{-6mm}}l}
\hline
$E_{\rm beam}$ & \multicolumn{3}{c}{$E^{\rm eff}_{\rm c.m.}$} & \multicolumn{3}{c}{Cross section} \\
 \cline{5-7}
{[MeV]}				&	 \multicolumn{3}{c}{[MeV]}		& \multicolumn{3}{c}{[mbarn]}		\\
\hline
13.0	&	12.62	&	$\pm$	&	0.04	&	0.139	&	$\pm$	&	0.016	\\
14.0	&	13.60	&	$\pm$	&	0.04	&	1.36	&	$\pm$	&	0.12	\\
15.0	&	14.57	&	$\pm$	&	0.05	&	8.03	&	$\pm$	&	0.73	\\
16.0	&	15.54	&	$\pm$	&	0.05	&	31.0	&	$\pm$	&	2.8	\\
17.0	&	16.52	&	$\pm$	&	0.05	&	98.3	&	$\pm$	&	9.0	\\
18.0	&	17.49	&	$\pm$	&	0.06	&	188	&	$\pm$	&	17	\\
18.9$^*$	&	18.33	&	$\pm$	&	0.11	&	281	&	$\pm$	&	25	\\
20.0	&	19.43	&	$\pm$	&	0.06	&	417	&	$\pm$	&	36	\\
\hline
\end{tabular*}
\vspace{-4mm}
\flushleft $^*$Measured with energy degrader foil.
\end{table}

\section{Theoretical analysis}
\label{sec:theo}

\subsection{General Remarks}
\label{sec:general_an}
It is the aim of the present study to provide further constraints for the
calculation of \al -induced cross sections on \smiv . These constraints will
be based on the new \smiv \ran \gdvii\ data from this work which will be
complemented by further information from experiments on \smiv \rag \gdviii\
capture \cite{Somorjai1998,Scholz2020} and \smiv \raa \smiv\ elastic
scattering \cite{Kiss2022,Mohr1997}.

The calculations are based on the statistical model approach. The following
calculations were made with the widely used computer code TALYS
\cite{TALYS}. In a schematic notation, the cross section of an \raX\ reaction
is given by: 
\begin{equation}
\sigma(\alpha,X) \sim \frac{T_{\alpha,0} T_X}{\sum_i T_i} = T_{\alpha,0}
\times b_X ,
\label{eq:StM}
\end{equation}
with the transmission coefficients $T_i$ into the $i$-th open channel and the
branching ratio $b_X = T_X / \sum_i T_i$ for the decay into the channel
$X$. Details of the formalism are given e.g.\ in
\cite{Rauscher_IJMPE2011}.

The $T_i$ are calculated from global optical potentials for the particle
channels and from the $\gamma$-ray strength function (GSF) for the photon
channel. Besides the explicit dependence of the $T_i$ on the optical
potentials and the GSF, the $T_i$ implicitly depend on the nuclear level
densities (LDs) of the respective residual nuclei because each $T_i$ is
composed of the sum over few low-lying states $j$ below the excitation energy
$E^\ast_0$ in the respective residual nucleus plus an integral over the LD for
excitation energies above $E^\ast_0$.

$T_{\alpha,0}$ represents the transmission in the entrance channel (with the
target \smiv\ in the ground state). Thus, $T_{\alpha,0}$ depends only on the
chosen \al -nucleus optical model potential (AOMP), but is independent of the
chosen LD and other parameters of the statistical model. $T_{\alpha,0}$ is the
most important quantity in Eq.~(\ref{eq:StM}) because it defines the total
reaction cross section, \stot, for \al\ + \smiv .

\begin{figure}[ht]
\includegraphics[width=0.9\columnwidth]{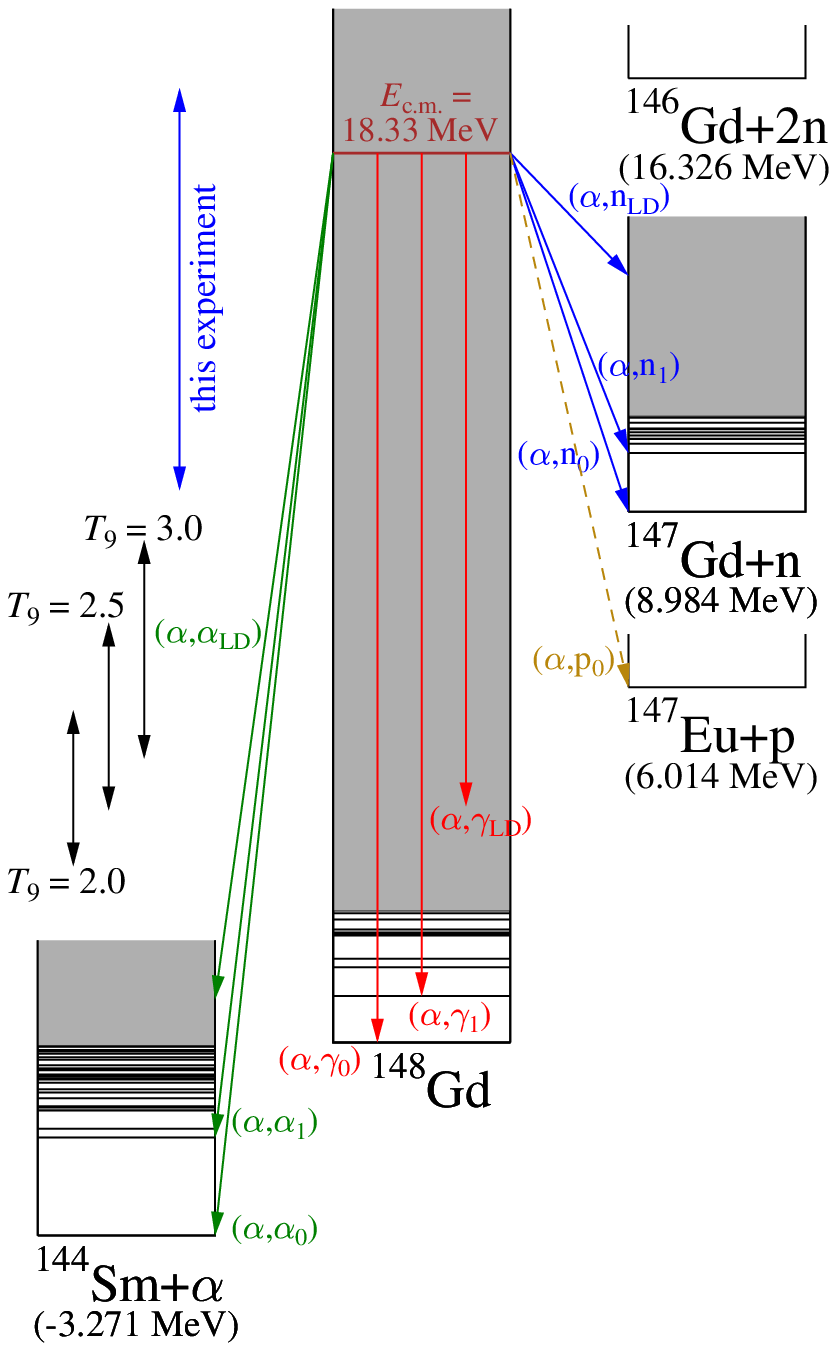}
\caption{
  \label{fig:level}
  Level scheme of \gdviii : The \smiv \raa \smiv , \smiv \rag \gdviii , \smiv
  \rap \euvii , \smiv \ran \gdvii , and \smiv \rann \gdvi\ reactions are
  illustrated. For each of the residual nuclei (\smiv , \gdviii , \euvii ,
  \gdvii , \gdvi ), the low-lying levels below $E_0^\ast$ are shown which are
  used explicitly in the statistical model calculations. The
  grey-shaded areas above $E_0^\ast$ are taken into account using the LD of
  the residual nuclei. As an example, the decay channels of the \gdviii\
  compound nucleus at $E_{\rm{c.m.}} = 18.33$ MeV (corresponding to an
  excitation energy $E^\ast = 15.06$ MeV in \gdviii ) are shown. For each
  \raX\ channel, the \raXnull\ decay to the ground state, \raXi\ decay to the
  first excited state, and one arrow for the \raXLD\ decay to higher-lying
  states above $E_0^\ast$ are shown; (\al ,$X_{i \ge 2}$) decays are not shown
  for better readability. The vertical arrows indicate the standard Gamow
  window of the \rag\ reaction for typical temperatures of $T = 2 - 3$
  GK ($T_9 = 2 - 3$). For further discussion see text.
}
\end{figure}

\begin{figure}
\includegraphics[width=\columnwidth]{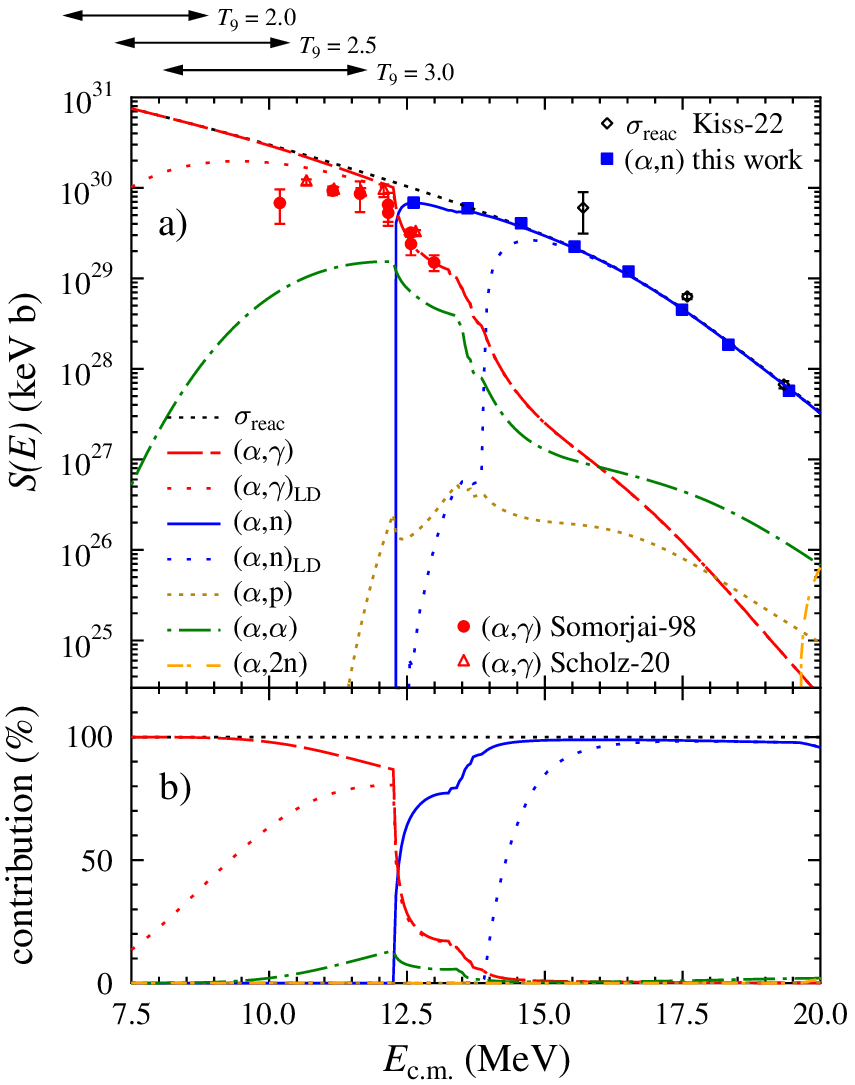}
\caption{
  \label{fig:decomp}
  Decomposition of the total reaction cross section, \stot, into the different
  \raX\ exit channels. All cross sections have been converted to astrophysical
  \sfact s in the upper part a); the lower part b) shows the contributions 
	of the different reaction channels to
  the total reaction cross section, \stot, in a linear scale. \raX $_{\rm{LD}}$
  shows the contribution of the higher-lying states above $E_0^\ast$ for the
  two dominating \raX\ channels which are the \rag\ and the \ran\ channel. The
  experimental data points represent total cross sections from \raa\ elastic
  scattering (black open diamonds, \cite{Kiss2022}), \ran\ cross sections from
  this work (blue full squares), and \rag\ cross sections (red full circles
  \cite{Somorjai1998} and red open triangles \cite{Scholz2020}).  The arrows
  indicate the standard Gamow window for typical temperatures of $T_9 = 2 - 3$. 
	For further discussion see text.
}
\end{figure}

From Eq.~(\ref{eq:StM}) several basic properties of the reactions under study
can be deduced. This is also illustrated in the level scheme
(Fig.~\ref{fig:level}) and in the decomposition of the total cross section, 
\stot, into the the different exit channels (Fig.~\ref{fig:decomp}).

In general, for heavy target nuclei and low energies, the transmission of the
\al -particle, $T_{\alpha,0}$, in the entrance channel is much smaller than
other $T_X$ like $T_\gamma$ or $T_n$. Thus, below the neutron threshold
$b_\gamma \approx 1$, and the \rag\ cross section depends only on
$T_{\alpha,0}$. Above the neutron threshold, $b_n \approx 1$, and the \ran\
cross section depends only on $T_{\alpha,0}$. As $T_{\alpha,0}$ depends only
on the chosen AOMP, experimental data for \rag\ cross sections below the
neutron threshold and for \ran\ cross sections above the neutron threshold are
appropriate to constrain the AOMP without ambiguities from other ingredients
of the statistical model. The relevance of the different exit channels is
shown in Fig.~\ref{fig:decomp}; here the Atomki-V2 potential
\cite{Mohr_PRL2020,Mohr_ADNDT2021} was used as
AOMP. Further discussion of the AOMP will be given later.

The above Eq.~(\ref{eq:StM}) is valid for laboratory experiments where the
target nucleus is in its ground state. Thus, the transmission in the entrance
channel is given by $T_{\alpha,0}$. Contrary, the high temperatures in the
stellar interior lead to thermal population of excited states in the target
nucleus.
As a
consequence, under stellar conditions Eq.~(\ref{eq:StM}) has to be extended to
take into account that excited states in the target nucleus are thermally
populated, leading to an entrance channel transmission $T_\alpha$ instead of
$T_{\alpha,0}$. But the transmissions $T_{\alpha,i>0}$ to excited states of
the target nucleus are typically smaller than $T_{\alpha,0}$ to the ground
state. Thus, $T_\alpha$ remains smaller than $T_\gamma$ or $T_n$ under typical
stellar conditions because the Coulomb barrier suppresses transitions to
excited states. Hence, stellar cross sections and the resulting
stellar reaction rates of \al -induced reactions depend mainly on the chosen
AOMP.

Furthermore, because $T_{\alpha,0}$ is the dominating contributor to the
total \al\ transmission, $T_\alpha$, the stellar reaction rate, \Nsv $^{\ast}$,
of the \rag\ reaction remains close to the laboratory rate, \Nsv $_{\rm{lab}}$,
which is calculated from the \rag\ cross section under laboratory conditions
(i.e., with \smiv\ in its $0^+$ ground state). Contrary, excited states in
\gdviii\ play an essential role for the stellar \Nsv $^\ast$. Consequently,
\rag\ experiments in the laboratory are well-suited to determine the stellar
\rag\ rate. The stellar \rga\ rate is then derived from the stellar \rag\ rate
using the formalism of detailed balance. For completeness we note that \rga\
experiments cannot provide the stellar \rga\ rate because of the missing
contributions of thermally excited states in the target nucleus.

\subsection{Relevance of different exit channels and sensitivities}
\label{sec:exit}
A detailed investigation of Fig.~\ref{fig:decomp} provides an excellent
intuitive way to understand the sensitivities on the chosen parameters for the
statistical model. It is obvious from Fig.~\ref{fig:decomp} that there are two
dominating channels. The \rag\ channel below the \ran\ threshold at
12.255 MeV, and the \ran\ channel above the \ran\ threshold. Compound-elastic
and compound-inelastic scattering reach a maximum contribution of about 15\%
close below the \ran\ threshold, but remain very marginal for most energies
(e.g., far less than 10\% below 10 MeV and above 13 MeV). The \rann\ channel
opens above the experimental energy range of this work at 19.6 MeV. Although
the \rann\ contribution increases steeply with energy, it remains below a few
per cent up to 20 MeV in Fig.~\ref{fig:decomp}. The contribution of the \rap\
channel remains far below 1\% in the energy range of
Fig.~\ref{fig:decomp}. This leads to the following approximate simplifications
of Eq.~(\ref{eq:StM}):

\rag\ cross section below the \ran\ threshold:
\begin{eqnarray}
\sigma(\alpha,\gamma) & \sim & T_{\alpha,0} \times \frac{T_{\gamma}}{T_\gamma +
                               T_\alpha}
\label{eq:StM_ag1}
\\
  & \approx & T_{\alpha,0} \quad \quad \quad \quad {\textrm{for}} \, E
              \lesssim 10 \, {\textrm{MeV}},
\label{eq:StM_ag1_approx}
\end{eqnarray}
  
\rag\ cross section above the \ran\ threshold:
\begin{eqnarray}
\sigma(\alpha,\gamma) & \sim & T_{\alpha,0} \times \frac{T_{\gamma}}{T_\gamma +
                               T_\alpha + T_n}
\label{eq:StM_ag2}
\\
  & \approx & T_{\alpha,0} \times \frac{T_\gamma}{T_n}\quad \quad
              {\textrm{for}} \, E \gtrsim 13 \, {\textrm{MeV}},
\label{eq:StM_ag2_approx}
\end{eqnarray}

\ran\ cross section above the \ran\ threshold:
\begin{eqnarray}
\sigma(\alpha,n) & \sim & T_{\alpha,0} \times \frac{T_n}{T_\gamma +
                          T_\alpha + T_n}
\label{eq:StM_an} \\
  & \approx & T_{\alpha,0} \quad \quad \quad \quad {\textrm{for}} \, E
              \gtrsim 13 \, {\textrm{MeV}}.
\label{eq:StM_an_approx}
\end{eqnarray}
  
All $T_i$ (except $T_{\alpha,0}$ in the entrance channel) are composed of a
sum over low-lying states and an integral over the level density in the
respective residual nucleus. By default, TALYS considers the first 10
low-lying excited states explicitly; the contributions of higher-lying excited
states are calculated using a continuous distribution of levels from a level
density formula. Thus, in principle all $T_i$ in
Eqs.~(\ref{eq:StM_ag1})$-$(\ref{eq:StM_an_approx}) and the resulting \rag\ and
\ran\ cross sections depend implicitly on the chosen level density. 

In practice, the dependence of the \rag\ and \ran\ cross sections on the
chosen level density remains marginal. In the case of the \ran\ cross section,
the low-lying states dominate only from threshold up to about 14 MeV; above 15
MeV, the contribution of the higher-lying excited states which is simulated by
the level density, exceeds 80\%. Any change in the level density affects
whether the \ran\ cross section is dominated by low-lying excited states or by
high-lying excited states, but does not affect the total \ran\ cross section
(as measured in the activation experiment). In the case of the \rag\ cross
section, the contribution of high-lying states is increasing with energy and
exceeds 50\% around 9 MeV; i.e., all available experimental data points of
\cite{Somorjai1998,Scholz2020} are governed by transitions to higher-lying
excited states in \gdviii . Nevertheless, a significant relevance of the level
density appears only above the \ran\ threshold, see also
Eq.~(\ref{eq:StM_ag2_approx}). Below the \ran\ threshold, the \rag\ cross
section practically depends only on $T_{\alpha,0}$ and thus on the AOMP, see
Eq.~(\ref{eq:StM_ag1_approx}).

Summarizing the above findings, the new \ran\ data can be used to constrain
the AOMP. The few available \rag\ data points above the \ran\ threshold
provide some information on the \g -ray strength function and the level
density of \gdviii , and the \rag\ data below the \ran\ threshold should be
well predicted because these \rag\ data depend only on the AOMP which is
well-constrained by the \ran\ data. Consequently, the following discussion
focuses on the AOMP whereas the other ingredients of the statistical model
calculations are only briefly mentioned. For completeness, we point out that
the \rag\ reaction rate for typical \gpro\ temperatures of $T_9 = 2 - 3$ 
is defined by the \rag\ cross section below the \ran\
threshold, and thus the \rag\ reaction rate also depends only on the AOMP.

\subsection{Comparison of experimental and theoretical \ran\ cross sections}
\label{sec:comp_an}
\begin{figure}
\includegraphics[width=\columnwidth,clip=]{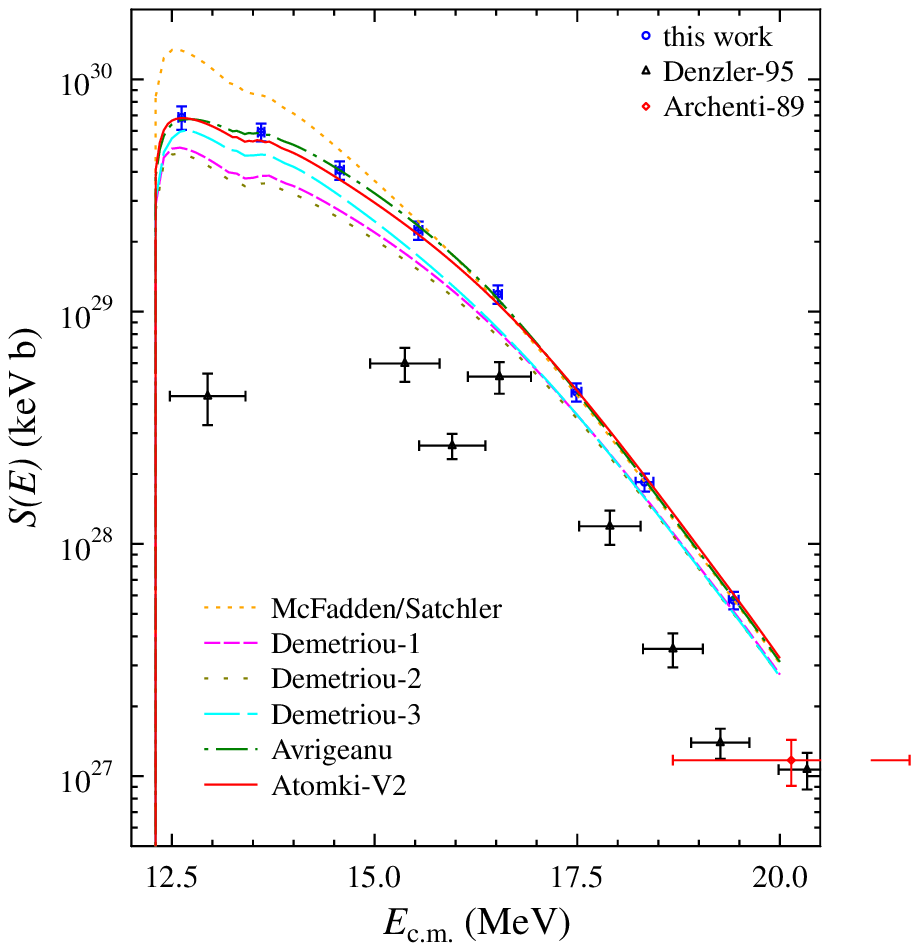}
\caption{
  \label{fig:an_aomp}
  Comparison of experimental \smiv \ran \gdvii\ cross section (shown as
  astrophysical \sfact ) to predictions from different AOMPs. Additional
  experimental data are taken from literature \cite{Denzler1995,Archenti1989};
  the lower data point of \cite{Archenti1989} at 13.1 MeV is about three
  orders of magnitude above the present data (not shown). The Atomki-V2 AOMP
  and the AOMP by Avrigeanu {\it et al.}\ \cite{Avrigeanu_PRC2014_aomp}
  reproduce all new experimental data within their error bars. 
	For further discussion see text.
}
\end{figure}

Fig.~\ref{fig:an_aomp} compares the new experimental \ran\ data to the
predictions of various AOMPs. It is obvious from Fig.~\ref{fig:an_aomp} that
the earlier data by Denzler {\it et al.}\ \cite{Denzler1995} and Archenti {\it
  et al.}\ \cite{Archenti1989} are not suitable to constrain the AOMP. In both
experiments, the stacked-foil technique was used which leads to considerable
uncertainties in the energy. In particular, when using degrader foils where uncertainties
in target foils and degrader foils sum up at the last targets. 
(Note that the lowest data point by Archenti {\it et al.}\ at 13.1 MeV is
not shown in Fig.~\ref{fig:an_aomp} because it is located about three orders
of magnitude above the present data and all calculations and is thus far out
of the scale of Fig.~\ref{fig:an_aomp}.)

The widely used simple 4-parameter AOMP by
McFadden and Satchler \cite{McFadden_NPA1966_aomp} fits the new data at higher
energies; however, at lower energies close above the threshold this AOMP
clearly overestimates the experimental data. Such an overestimation towards
lower energies has been found also for the \smiv \rag \gdviii\ reaction
\cite{Somorjai1998,Scholz2020}. An explanation for the overestimation of
low-energy cross sections was provided in \cite{Mohr_PRL2020} which is related
to the tail of the imaginary potential at large radii above 10 fm (far beyond
the colliding nuclei).

The three versions of the AOMP by Demetriou {\it et al.}\
\cite{Demetriou_NPA2002_aomp} underestimate the new \ran\ data over the whole
energy range under study. The most elaborated version 3 of the Demetriou
potentials is closer to the experimental data than versions 1 and 2
which do not take into account the dispersive coupling between the real and
imaginary parts of the AOMP.

The Atomki-V2 AOMP \cite{Mohr_PRL2020} and the AOMP by Avrigeanu {\it et al.}\
\cite{Avrigeanu_PRC2014_aomp} reproduce all new \ran\ data points within their
small experimental uncertainties, which is a remarkable success for both
AOMPs. As a consequence, the new \ran\ data can recommend these two AOMPs, but
cannot provide a preferred AOMP for the calculation of the \smiv \rag \gdviii\
reaction rate.

It has been shown in earlier work (e.g.,
\cite{Szucs_PRC2019_au197aX,Mohr_PRC2018_38ar,Mohr_PRC2017_64zn}) that three
further AOMPs which are available in TALYS, are not recommended for the
calculation of low-energy cross sections. These potentials by Watanabe
\cite{Watanabe_NPA1958_omp}, Nolte {\it et al.}\ \cite{Nolte_PRC1987_aomp}, and
Avrigeanu {\it et al.}\ \cite{Avrigeanu_PRC1994_aomp} are omitted in the
present analysis.

\subsection{Discussion of the various $\alpha$-nucleus potentials}
\label{sec:disc_aomp}
In this section we provide some basic information on the various AOMPs under
study in the present work which are the AOMPs by McFadden and Satchler
\cite{McFadden_NPA1966_aomp} (MCF), Demetriou {\it et al.}\
\cite{Demetriou_NPA2002_aomp} (third version of these AOMPs, DEM-3), Avrigeanu
{\it et al.}\ \cite{Avrigeanu_PRC2014_aomp} (AVR), and the new Atomki-V2 AOMP
\cite{Mohr_PRL2020}.

The MCF AOMP was derived from elastic scattering at energies around 25 MeV. It
is a very simple energy-independent 4-parameter AOMP. Because of its
simplicity, the MCF AOMP is widely used; e.g., the NON-SMOKER calculations of
reaction rates \cite{Rauscher_ADNDT2000_rates} utilize the MCF AOMP, and
these rates are adopted in the REACLIB database used in many astrophysical models
\cite{Cyburt_APJS2010_reaclib}. Whereas the MCF AOMP can be applied
successfully to low-energy data in the $A \approx 20 - 50$ mass range
\cite{Mohr_EPJA2015_A20-50}, the MCF AOMP fails at sub-Coulomb energies in the
heavy mass range; see, e.g., \cite{Somorjai1998,Scholz2020} for the \smiv \rag
\gdviii\ reaction and
\cite{Szucs_PRC2019_au197aX,Sauerwein_PRC2011_141pr,Mohr_PRC2011_pr141a} for
other target nuclei.

The DEM-3 AOMP has become the basis for \al -induced reaction rates in the
STARLIB database \cite{Sallaska_APJS2013_starlib}. The DEM-3 AOMP is based on
the double-folding approach and takes into account the dispersive coupling
between the real and imaginary parts of the AOMP. Its parameters were adjusted
to a limited set of low-energy reaction and scattering data which was
available about 20 years ago. In particular, this data set includes the \smiv
\rag \gdviii\ data of Somorjai {\it et al.}\ \cite{Somorjai1998}. Because of
the high astrophysical relevance, much effort was spent for the DEM-3 AOMP to
fit the \rag\ data point at the lowest energy which shows an unexpectedly
small cross section with a huge error bar. This attempt to fit the very small
\rag\ cross section may be one explanation for the underestimation of the
\ran\ cross sections of the present work using the DEM-3 AOMP.

Similar to the DEM-3 AOMP, the AVR AOMP is also based on the double-folding
approach. However, at the end a Woods-Saxon parametrization was introduced
\cite{Avrigeanu_NPA2003_aomp}, and the Woods-Saxon parameters were fine-tuned
to a wider data set of reaction cross sections and elastic scattering angular
distributions, finally leading to an energy-dependent many-parameter AOMP which
was able to reproduce practically all available experimental data at that
time. Because of this success, the AVR AOMP has been adopted as the default
AOMP in TALYS for several years.

Similar to the DEM-3 and AVR AOMPs, the Atomki-V2 AOMP is based on the
double-folding approach. Contrary to the previous AOMPs, its parameters are
obtained from a compilation of low-energy elastic scattering only
\cite{Mohr_ADNDT2013_atomki-v1}; no adjustment to reaction data was
made. As it was noticed that usual optical model calculations become
extremely sensitive to the tail of the imaginary potential at extreme
sub-Coulomb energies, the Atomki-V2 AOMP uses a very narrow, deep, and
sharp-edged imaginary part which avoids complications with the tail of the
imaginary potential and leads to cross sections similar to a simple barrier
transmission approach. It is interesting to note that this simple barrier
transmission approach in combination with the energy-independent Atomki-V2
AOMP is able to predict reaction cross sections without any adjustment to
reaction data; these predictions are for many targets surprisingly close to
the energy-dependent multi-parameter approach of the AVR AOMP. This holds also
for the present study of the \smiv \ran \gdvii\ reaction (see
Fig.~\ref{fig:an_aomp}).

\section{Astrophysical reaction rate of the $^{144}$S\lowercase{m}($\alpha,\gamma$)$^{148}$G\lowercase{d} reaction}
\label{sec:ag_rate}
\subsection{General Remarks}
\label{sec:general_ag}
The \smiv \rag \gdviii\ reaction has found much attention over the previous
decades because its inverse \gdviii \rga \smiv\ reaction has strong a impact on
the production ratio, $P_{146/144}$, for the two samarium isotopes \smiv\ and
\smvi\ under \gpro\ conditions. \smiv\ is a stable \pnuc\ which is directly
produced by \gdviii \rga \smiv, whereas \smvi\ is an unstable \al -emitter
with a half-life of about 100 million years which is produced by \gdviii \rgn
\gdvii \rgn \gdvi\ and subsequent $\beta$-decays to $^{146}$Eu and \smvi
. Nowadays, an excess of the $^{142}$Nd/$^{144}$Nd ratio is found in
correlation with the samarium-to-neodymium ratio in meteoritic samples. This
excess of $^{142}$Nd reflects the \al -decay of \smvi\ and can be used as a
cosmochronometer, provided that the initial production ratio, $P_{146/144}$, can
be calculated reliably and the ratio at the formation of our solar system
$P_{146/144} = 0.00840(32)$ \cite{LinruFang2022} is well-known, see e.g.\ 
\cite{Rauscher2013,LinruFang2022,Lugaro2016_PNAS,Rauscher2013_PRL,Kinoshita2012_Science,Gannoun2011}
and references therein, including the basic ideas in
\cite{Woosley_APJS1978,Audouze1972} and early measurements
\cite{Prinzhofer1989,Prinzhofer1992,Nyquist1994,Rankenburg2006}.

\subsection{Comparison of experimental and theoretical \rag\ cross sections}
\label{sec:comp_ag}
\begin{figure}
\includegraphics[width=\columnwidth,clip=]{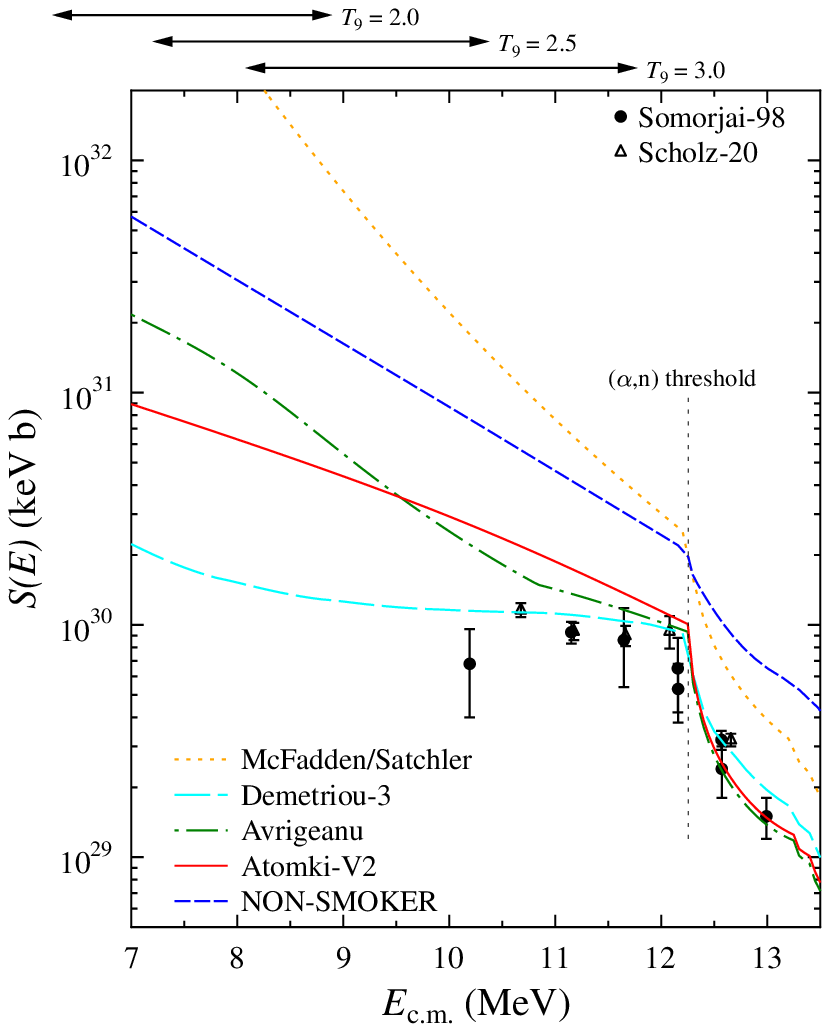}
\caption{
  \label{fig:ag_aomp}
  Comparison of experimental \smiv \rag \gdviii\ cross section (shown as
  astrophysical \sfact ) to predictions from different AOMPs. The experimental
  data are taken from \cite{Somorjai1998,Scholz2020}. The thin vertical line
  indicates the \ran\ threshold. For energies below, the \rag\ cross section
  depends practically only on the AOMP. The horizontal bars on top indicate
  the standard Gamow windows for typical temperatures of the \gpro . Further
  discussion see text.
}
\end{figure}

Fig.~\ref{fig:ag_aomp} compares the experimental data of the \smiv \rag
\gdviii\ reaction to the predictions of selected AOMPs (see previous
Sect.~\ref{sec:disc_aomp}). As the \rag\ cross section below the \ran\
threshold depends practically only on the chosen AOMP, one should expect that
the AVR and Atomki-V2 AOMPs are able to reproduce the experimental \rag\ data
because the AVR and Atomki-V2 AOMPs worked excellent for the \ran\ data (see
Fig.~\ref{fig:an_aomp}). Somewhat surprisingly, this is not the case. The
predictions from the AVR and Atomki-V2 AOMPs remain close to each other within
about $20 - 30$ per cent above 9 MeV. Only towards very low energies below 8
MeV, the AVR prediction exceeds the Atomki-V2 prediction by more than a
factor of two.

Both the AVR and Atomki-V2 predictions are, however, a factor of about $1.5 - 2$ above
the low-energy data of \cite{Somorjai1998,Scholz2020}. Contrary, the DEM-3
AOMP fits the experimental \rag\ data quite well, but was not able to fit the
new \ran\ data. A simultaneous description of the \rag\ and \ran\ data is not
possible within the available AOMPs and would require to introduce a very
special energy dependence. However, such a special energy dependence may lead
to significant uncertainties in the extrapolation towards energies below the
lowest experimental \rag\ data points. Such an extrapolation is necessary for
the calculation of the astrophysical reaction rate \Nsv .

As expected, the MCF AOMP overestimates the experimental \rag\ cross sections
at all energies. This overestimation becomes most pronounced towards low
energies. Somewhat surprisingly, the \rag\ cross sections from the NON-SMOKER
code \cite{NONSMOKER} deviate significantly from the present calculation using
the MCF AOMP. As also NON-SMOKER uses the MCF AOMP, this difference must result
from different numerical treatments in the TALYS and NON-SMOKER codes.

\subsection{Recommendations for the \rag\ reaction rate}
\label{sec:ag_rate_rec}
Taking into account the main result of the previous Sect.~\ref{sec:comp_ag}
that it is practically impossible to fit simultaneously the new \ran\ data and
the literature \rag\ data, it remains a difficult task to provide a reliable
reaction rate, \Nsv, of the \smiv \rag \gdviii\ reaction. Despite these
problems, significant progress is achieved when compared to the
widely used conclusion that \rag\ rates are uncertain by at least a factor of
10 (see e.g.\ \cite{Rauscher_MNRAS2016}).

\begin{figure}
\includegraphics[width=\columnwidth,clip=]{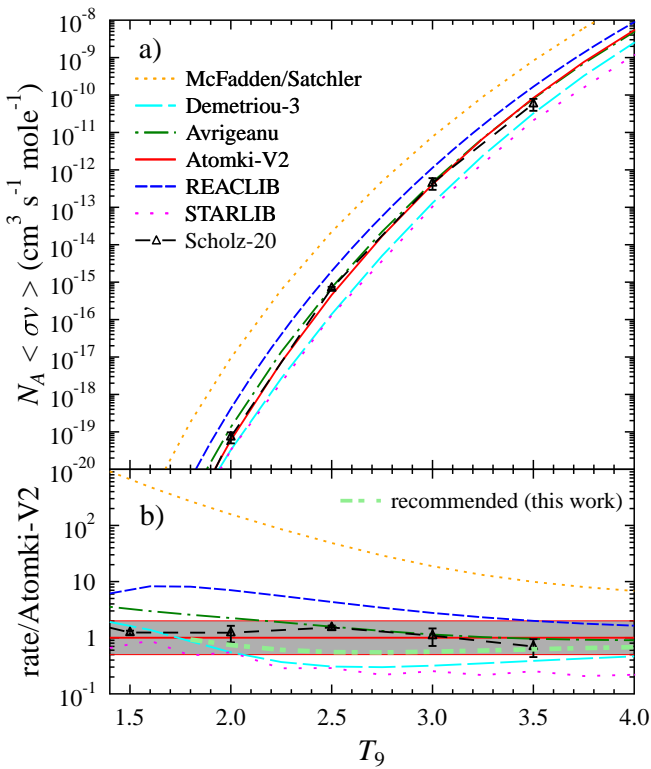}
\caption{
  \label{fig:ag_rate}
  Astrophysical reaction rate, \Nsv, from different AOMPs (upper part a) and
  ratio to the rate from the Atomki-V2 AOMP (lower part b). The gray-shaded
  band shows the expected uncertainty of a factor of two from the Atomki-V2
  AOMP \cite{Mohr_PRL2020}. For further discussion see text.
}
\end{figure}

We have calculated the astrophysical reaction rate, \Nsv, for the different
AOMPs under study. The results are shown in Fig.~\ref{fig:ag_rate}. For the
interpretation of the \rag\ cross sections in Fig.~\ref{fig:ag_aomp} and the
reaction rates, \Nsv, in Fig.~\ref{fig:ag_rate} it is important to note that
the classical Gamow windows (as e.g.\ indicated by the horizontal arrows on
top of Fig.~\ref{fig:ag_aomp}) are calculated under the assumption of a
constant (energy-independent) astrophysical \sfact . However, for heavy nuclei
the \sfact\ typically has a noticeable negative slope, leading to a shift of
the real Gamow window towards lower energies by about 1 MeV
\cite{Rauscher_PRC2010_gamow}. Because the reaction rate, \Nsv, in the upper
part of Fig.~\ref{fig:ag_rate} is an extremely temperature-dependent quantity,
for better comparison the lower part of Fig.~\ref{fig:ag_rate} shows the ratio
of the respective rates to the rate from the Atomki-V2 AOMP.

It was shown in \cite{Mohr_PRL2020} that the predictions from the Atomki-V2
AOMP typically match experimental data with deviations below a factor of two
for a wide range of heavy target nuclei. Consequently, astrophysical reaction
rates from the Atomki-V2 AOMP should be well-defined within an uncertainty
factor of two. This assumed uncertainty is indicated by the grey-shaded error
band in Fig.~\ref{fig:ag_rate}.

Because of the close similarity of the calculated \rag\ cross sections from
the Atomki-V2 and AVR AOMPs, the rate from the AVR AOMP remains well within the
estimated uncertainty band from the Atomki-V2 AOMP for typical temperatures of
the \gpro\ of about $T_9 = 2 - 3$. Only towards lower temperatures, the rate
from the AVR AOMP clearly exceeds the rate from the Atomki-V2 AOMP; this
excess corresponds to the steeply increasing \sfact\ of the AVR AOMP towards
lower energies. Around 7 MeV the cross section from the AVR AOMP is higher by
a factor of two than the Atomki-V2 cross section, leading to an enhanced rate
by a factor of two at $T_9 \approx 2$.

The rate from the DEM-3 AOMP is lower by about a factor of $2-3$ in the
temperature range of the \gpro . However, at very low temperatures below
$T_9 \approx 1.7$ the rate from the DEM-3 AOMP exceeds the Atomki-V2 rate;
this results from a steep increase of the astrophysical \sfact\ for the DEM-3
AOMP at energies below the shown range of Fig.~\ref{fig:ag_aomp}. The STARLIB
rate which is based on the DEM-3 AOMP is close to the present calculation of
the DEM-3 AOMP. The minor differences may arise from the earlier TALYS version
used for the STARLIB rates and from the enhanced accuracy settings in the
present rate calculation.

The REACLIB rate is based on the NON-SMOKER calculation of the \rag\ cross
section using the MCF AOMP (see also Fig.~\ref{fig:ag_aomp} and discussion
above). As this cross section does not agree with the present calculation
using the MCF AOMP, it is not recommended to use the REACLIB rate for the
\smiv \rag \gdviii\ reaction.

The rate by Scholz {\it et al.}\ is based on a hyperparameter optimization of
TALYS calculations which were adjusted to the available experimental \smiv
\rag \gdviii\ data \cite{Scholz2020,Somorjai1998}. This rate is remarkably
close to the rates from the Atomki-V2 and AVR AOMPs.

Before a final recommendation of the \smiv \rag \gdviii\ rate can be given,
the apparent tension between the lowest experimental \rag\ data points by
Somorjai {\it et al.}\ \cite{Somorjai1998} and Scholz {\it et al.}\
\cite{Scholz2020} (see also Fig.~\ref{fig:ag_aomp}) needs further
investigation. For shorter notation, all \sfact s in the following paragraphs
are given in units of $10^{30}$ keV\,b.

Somorjai {\it el al.}\ report a \sfact\ of 0.713(292) at $E = 10.193$
MeV. Using an updated half-life of \gdviii , Scholz {\it et al.}\ correct that
value to 0.68(28). For a consistent comparison of the two activation data
sets, the same half-life of \gdviii\ has to be used; thus, the latter value is
taken for comparison here. Scholz {\it et al.}\ report their lowest \sfact\ of
1.16(8) at $E = 10.675$ MeV.

For a comparison at $E = 10.193$ MeV, the \sfact\ by Scholz {\it et al.}\ at
$10.675$ MeV has to be extrapolated down to $E = 10.193$ MeV. For this
purpose, the theoretical energy dependence from the recent AOMPs was used. The
ratio between the \sfact s at 10.193 MeV and 10.675 MeV is 1.03 (1.35, 1.24)
from the DEM-3 (AVR, Atomki-V2) AOMPs; the average ratio is 1.21. This number
changes only marginally to 1.26 if further AOMPs (DEM-1, DEM-2, MCF) are considered. 
This leads to an extrapolated \sfact\ at 10.193 MeV of 1.40 and
an interval between 1.20 and 1.57 from the lowest and highest
theoretical ratios. Thus, the extrapolation procedure leads to an additional
uncertainty of about 14\% for the \sfact\ at 10.193 MeV which has to be added
to the 7\% uncertainty of the original data point of Scholz {\it et al.}

The \sfact\ and uncertainty of 0.68(28) of Somorjai {\it et al.}\ at 10.193
MeV may be slightly misleading; this becomes obvious from the resulting
3-$\sigma$ interval from $-0.16$ to $1.52$ with its non-physical negative
lower limit. Relatively large uncertainties should be provided as an
uncertainty factor of the underlying log-normal distribution. A careful
estimate from the lower 1-$\sigma$ interval leads to an uncertainty factor of
$\frac{0.68}{0.68-0.28} = 1.70$ for the lowest data point of Somorjai {\it et
  al.}  Using this uncertainty factor of 1.70 for a realistic estimate of the
upper end of the 1-$\sigma$ interval leads to an upper \sfact\ of 1.16 which is
quite close to the lower 1-$\sigma$ limit of the extrapolated data point by
Scholz {\it et al.}  (as derived in the previous paragraph). Consequently, the
apparent tension between the lowest data points of Somorjai {\it et al.}\ and
Scholz {\it et al.}\ is not very significant and results mainly from the
misunderstandable linear specification of the uncertainties by Somorjai {\it
  et al.}

This leads us finally to the following recommendation for the \smiv \rag
\gdviii\ reaction rate. The rate from the Atomki-V2 AOMP should be considered
as an upper limit because Atomki-V2 fits the \ran\ data very well, but
overestimates the \rag\ data at low energies. Contrary, the DEM-3 AOMP fits
the \rag\ data well, but systematically underestimates the \ran\ data. Thus,
the rate from the DEM-3 AOMP can be considered as a lower limit of the
rate. This consideration is further strengthened by the relatively small slope
of the astrophysical \sfact\ from the DEM-3 AOMP towards lower energies down
to about 7 MeV. The disagreement between the DEM-3 \sfact\ and the lowest
experimental data point by Somorjai {\it et al.}\ is not very significant
because of the large uncertainty of this data point (see discussion in the
previous paragraphs). The recommended rate is then derived from the geometric
mean of the lower and upper limits of the rate. The results are listed in
Table \ref{tab:rate}.
\begin{table}
  \caption{
    \label{tab:rate}
    Recommended reaction rate, \Nsv, and lower and upper limits for the \smiv
    \rag \gdviii\ reaction. All rates are given in units of cm$^3$ s$^{-1}$
    mole$^{-1}$.
  }
\begin{tabular}{cr@{$\times 10$}lr@{$\times 10$}lr@{$\times 10$}l}
\hline
  $T_9$
  & \multicolumn{2}{c}{lower limit}
  & \multicolumn{2}{c}{upper limit}
  & \multicolumn{2}{c}{recommended} \\
\hline
  2.00 & 3.29 & $^{-20}$ & 5.97 & $^{-20}$ & 4.43 & $^{-20}$ \\
  2.25 & 2.68 & $^{-18}$ & 7.27 & $^{-18}$ & 4.42 & $^{-18}$ \\
  2.50 & 1.39 & $^{-16}$ & 4.55 & $^{-16}$ & 2.52 & $^{-16}$ \\
  2.75 & 5.06 & $^{-15}$ & 1.69 & $^{-14}$ & 9.26 & $^{-15}$ \\
  3.00 & 1.31 & $^{-13}$ & 4.11 & $^{-13}$ & 2.32 & $^{-13}$ \\
  3.50 & 3.23 & $^{-11}$ & 8.36 & $^{-11}$ & 5.20 & $^{-11}$ \\
  4.00 & 2.51 & $^{-09}$ & 5.40 & $^{-09}$ & 3.68 & $^{-09}$ \\
\hline
\end{tabular}
\end{table}

It is important to point out that the upper and lower limits of the rate do
not deviate by more than a factor of $3-4$ over the whole relevant temperature
range of $T_9 = 2 - 3$. Thus, the recommended rate does not deviate by more
than a factor of two from the Atomki-V2 rate and the DEM-3 rate. A slightly
larger deviation of a factor of about 3 is found between the recommended rate
and the rates from the AVR AOMP and from the hyperparameter optimization in
\cite{Scholz2020}. Compared to earlier conclusions that the reaction rates of
\al -induced reactions are very uncertain by at least a factor of 10, the
situation has improved significantly for the \smiv \rag \gdviii\ reaction by
a major reduction of the uncertainty.

\section{Summary and conclusions}
\label{Sec:conc}
The \smiv \ran \gdvii\ reaction was studied at energies from the \ran\
threshold up to about 20 MeV. The new experimental data have significantly
smaller uncertainties than the few available data in literature
\cite{Denzler1995,Archenti1989}. Thus, the data can be used to constrain the
\al -nucleus optical model potential because a sensitivity study shows that
the new \ran\ data are almost exclusively sensitive to the chosen AOMP.

It is found that the new experimental data can be described very well using
the Atomki-V2 and AVR AOMPs. However, it turns out that these two AOMPs
slightly overestimate the existing data for \smiv \rag \gdviii \
\cite{Scholz2020,Somorjai1998}. This is a somewhat surprising result as the
\rag\ data are also sensitive only to the chosen AOMP. Contrary, the DEM-3
AOMP fits the \rag\ data better, but clearly underestimates the new \ran\ data.

As a consequence, our final recommended rate has to result from a compromise
between the lower rate from the DEM-3 AOMP and the higher rates from the
Atomki-V2 and AVR AOMPs. On the one hand, this is not a fully satisfying
situation because of the tension in the reproduction of the new \ran\ data and
the \rag\ data from literature which calls for further investigations. On the
other hand, the uncertainty of the recommended rate is now much smaller when
compared to earlier estimates of a factor of 10 or more. The achieved accuracy
of the \smiv \rag \gdviii\ rate in combination with the overall improvement
for \al -induced reaction rates \cite{Mohr_ADNDT2021} should now permit 
stronger constraints for astrophysical conclusions for the \gpro ,
similar to what has already been achieved very recently for the weak \rpro\
\cite{Psaltis2022}.

\begin{acknowledgments}
This work was supported by NKFIH grants No. NN128072 and K134197 and by the New National Excellence Programs of the Ministry of Human Capacities of Hungary (\'UNKP-22-5-DE-428). The financial support of the Hungarian Academy of Sciences (Infrastructure grants), and the Economic Development and Innovation Operational Programme (GINOP-2.3.3–15-2016-00005) grant, co-funded by the EU, is also acknowledged. T. Sz\"ucs acknowledges support from the Bolyai research fellowship. 
\end{acknowledgments}

%


\begin{thebibliography}{58}%
\makeatletter
\providecommand \@ifxundefined [1]{%
 \@ifx{#1\undefined}
}%
\providecommand \@ifnum [1]{%
 \ifnum #1\expandafter \@firstoftwo
 \else \expandafter \@secondoftwo
 \fi
}%
\providecommand \@ifx [1]{%
 \ifx #1\expandafter \@firstoftwo
 \else \expandafter \@secondoftwo
 \fi
}%
\providecommand \natexlab [1]{#1}%
\providecommand \enquote  [1]{``#1''}%
\providecommand \bibnamefont  [1]{#1}%
\providecommand \bibfnamefont [1]{#1}%
\providecommand \citenamefont [1]{#1}%
\providecommand \href@noop [0]{\@secondoftwo}%
\providecommand \href [0]{\begingroup \@sanitize@url \@href}%
\providecommand \@href[1]{\@@startlink{#1}\@@href}%
\providecommand \@@href[1]{\endgroup#1\@@endlink}%
\providecommand \@sanitize@url [0]{\catcode `\\12\catcode `\$12\catcode
  `\&12\catcode `\#12\catcode `\^12\catcode `\_12\catcode `\%12\relax}%
\providecommand \@@startlink[1]{}%
\providecommand \@@endlink[0]{}%
\providecommand \url  [0]{\begingroup\@sanitize@url \@url }%
\providecommand \@url [1]{\endgroup\@href {#1}{\urlprefix }}%
\providecommand \urlprefix  [0]{URL }%
\providecommand \Eprint [0]{\href }%
\providecommand \doibase [0]{https://doi.org/}%
\providecommand \selectlanguage [0]{\@gobble}%
\providecommand \bibinfo  [0]{\@secondoftwo}%
\providecommand \bibfield  [0]{\@secondoftwo}%
\providecommand \translation [1]{[#1]}%
\providecommand \BibitemOpen [0]{}%
\providecommand \bibitemStop [0]{}%
\providecommand \bibitemNoStop [0]{.\EOS\space}%
\providecommand \EOS [0]{\spacefactor3000\relax}%
\providecommand \BibitemShut  [1]{\csname bibitem#1\endcsname}%
\let\auto@bib@innerbib\@empty
\bibitem [{\citenamefont {Arnould}\ and\ \citenamefont
  {Goriely}(2003)}]{Arnould2003}%
  \BibitemOpen
  \bibfield  {author} {\bibinfo {author} {\bibfnamefont {M.}~\bibnamefont
  {Arnould}}\ and\ \bibinfo {author} {\bibfnamefont {S.}~\bibnamefont
  {Goriely}},\ }\href
  {https://doi.org/https://doi.org/10.1016/S0370-1573(03)00242-4} {\bibfield
  {journal} {\bibinfo  {journal} {Physics Reports}\ }\textbf {\bibinfo {volume}
  {384}},\ \bibinfo {pages} {1} (\bibinfo {year} {2003})}\BibitemShut {NoStop}%
\bibitem [{\citenamefont {Rauscher}\ \emph {et~al.}(2013)\citenamefont
  {Rauscher}, \citenamefont {Dauphas}, \citenamefont {Dillmann}, \citenamefont
  {Fr{\"o}hlich}, \citenamefont {F{\"u}l{\"o}p},\ and\ \citenamefont
  {Gy{\"u}rky}}]{Rauscher2013}%
  \BibitemOpen
  \bibfield  {author} {\bibinfo {author} {\bibfnamefont {T.}~\bibnamefont
  {Rauscher}}, \bibinfo {author} {\bibfnamefont {N.}~\bibnamefont {Dauphas}},
  \bibinfo {author} {\bibfnamefont {I.}~\bibnamefont {Dillmann}}, \bibinfo
  {author} {\bibfnamefont {C.}~\bibnamefont {Fr{\"o}hlich}}, \bibinfo {author}
  {\bibfnamefont {Z.}~\bibnamefont {F{\"u}l{\"o}p}},\ and\ \bibinfo {author}
  {\bibfnamefont {G.}~\bibnamefont {Gy{\"u}rky}},\ }\href
  {https://doi.org/10.1088/0034-4885/76/6/066201} {\bibfield  {journal}
  {\bibinfo  {journal} {Reports on Progress in Physics}\ }\textbf {\bibinfo
  {volume} {76}},\ \bibinfo {pages} {066201} (\bibinfo {year}
  {2013})}\BibitemShut {NoStop}%
\bibitem [{\citenamefont {Travaglio}\ \emph {et~al.}(2011)\citenamefont
  {Travaglio}, \citenamefont {R{\"o}pke}, \citenamefont {Gallino},\ and\
  \citenamefont {Hillebrandt}}]{Travaglio2011}%
  \BibitemOpen
  \bibfield  {author} {\bibinfo {author} {\bibfnamefont {C.}~\bibnamefont
  {Travaglio}}, \bibinfo {author} {\bibfnamefont {F.~K.}\ \bibnamefont
  {R{\"o}pke}}, \bibinfo {author} {\bibfnamefont {R.}~\bibnamefont {Gallino}},\
  and\ \bibinfo {author} {\bibfnamefont {W.}~\bibnamefont {Hillebrandt}},\
  }\href {https://doi.org/10.1088/0004-637X/739/2/93} {\bibfield  {journal}
  {\bibinfo  {journal} {The Astrophysical Journal}\ }\textbf {\bibinfo {volume}
  {739}},\ \bibinfo {pages} {93} (\bibinfo {year} {2011})}\BibitemShut
  {NoStop}%
\bibitem [{\citenamefont {K{\"a}ppeler}\ \emph {et~al.}(2011)\citenamefont
  {K{\"a}ppeler}, \citenamefont {Gallino}, \citenamefont {Bisterzo},\ and\
  \citenamefont {Aoki}}]{Kappeler2011}%
  \BibitemOpen
  \bibfield  {author} {\bibinfo {author} {\bibfnamefont {F.}~\bibnamefont
  {K{\"a}ppeler}}, \bibinfo {author} {\bibfnamefont {R.}~\bibnamefont
  {Gallino}}, \bibinfo {author} {\bibfnamefont {S.}~\bibnamefont {Bisterzo}},\
  and\ \bibinfo {author} {\bibfnamefont {W.}~\bibnamefont {Aoki}},\ }\href
  {https://doi.org/10.1103/RevModPhys.83.157} {\bibfield  {journal} {\bibinfo
  {journal} {Rev. Mod. Phys.}\ }\textbf {\bibinfo {volume} {83}},\ \bibinfo
  {pages} {157} (\bibinfo {year} {2011})}\BibitemShut {NoStop}%
\bibitem [{\citenamefont {Cowan}\ \emph {et~al.}(2021)\citenamefont {Cowan},
  \citenamefont {Sneden}, \citenamefont {Lawler}, \citenamefont {Aprahamian},
  \citenamefont {Wiescher}, \citenamefont {Langanke}, \citenamefont
  {Mart\'{\i}nez-Pinedo},\ and\ \citenamefont {Thielemann}}]{Cowan2021}%
  \BibitemOpen
  \bibfield  {author} {\bibinfo {author} {\bibfnamefont {J.~J.}\ \bibnamefont
  {Cowan}}, \bibinfo {author} {\bibfnamefont {C.}~\bibnamefont {Sneden}},
  \bibinfo {author} {\bibfnamefont {J.~E.}\ \bibnamefont {Lawler}}, \bibinfo
  {author} {\bibfnamefont {A.}~\bibnamefont {Aprahamian}}, \bibinfo {author}
  {\bibfnamefont {M.}~\bibnamefont {Wiescher}}, \bibinfo {author}
  {\bibfnamefont {K.}~\bibnamefont {Langanke}}, \bibinfo {author}
  {\bibfnamefont {G.}~\bibnamefont {Mart\'{\i}nez-Pinedo}},\ and\ \bibinfo
  {author} {\bibfnamefont {F.-K.}\ \bibnamefont {Thielemann}},\ }\href
  {https://doi.org/10.1103/RevModPhys.93.015002} {\bibfield  {journal}
  {\bibinfo  {journal} {Rev. Mod. Phys.}\ }\textbf {\bibinfo {volume} {93}},\
  \bibinfo {pages} {015002} (\bibinfo {year} {2021})}\BibitemShut {NoStop}%
\bibitem [{\citenamefont {Sz{\"u}cs}\ \emph {et~al.}(2019)\citenamefont
  {Sz{\"u}cs}, \citenamefont {Mohr}, \citenamefont {Gy{\"u}rky}, \citenamefont
  {Hal{\'a}sz}, \citenamefont {Husz{\'a}nk}, \citenamefont {Kiss},
  \citenamefont {Szegedi}, \citenamefont {T{\"o}r{\"o}k},\ and\ \citenamefont
  {F{\"u}l{\"o}p}}]{Szucs2019}%
  \BibitemOpen
  \bibfield  {author} {\bibinfo {author} {\bibfnamefont {T.}~\bibnamefont
  {Sz{\"u}cs}}, \bibinfo {author} {\bibfnamefont {P.}~\bibnamefont {Mohr}},
  \bibinfo {author} {\bibfnamefont {G.}~\bibnamefont {Gy{\"u}rky}}, \bibinfo
  {author} {\bibfnamefont {Z.}~\bibnamefont {Hal{\'a}sz}}, \bibinfo {author}
  {\bibfnamefont {R.}~\bibnamefont {Husz{\'a}nk}}, \bibinfo {author}
  {\bibfnamefont {G.~G.}\ \bibnamefont {Kiss}}, \bibinfo {author}
  {\bibfnamefont {T.~N.}\ \bibnamefont {Szegedi}}, \bibinfo {author}
  {\bibfnamefont {Z.}~\bibnamefont {T{\"o}r{\"o}k}},\ and\ \bibinfo {author}
  {\bibfnamefont {Z.}~\bibnamefont {F{\"u}l{\"o}p}},\ }\href
  {https://doi.org/10.1103/PhysRevC.100.065803} {\bibfield  {journal} {\bibinfo
   {journal} {Phys. Rev. C}\ }\textbf {\bibinfo {volume} {100}},\ \bibinfo
  {pages} {065803} (\bibinfo {year} {2019})}\BibitemShut {NoStop}%
\bibitem [{\citenamefont {Rauscher}(2011{\natexlab{a}})}]{Rauscher2011}%
  \BibitemOpen
  \bibfield  {author} {\bibinfo {author} {\bibfnamefont {T.}~\bibnamefont
  {Rauscher}},\ }\href {https://doi.org/10.1142/S021830131101840X} {\bibfield
  {journal} {\bibinfo  {journal} {International Journal of Modern Physics E}\
  }\textbf {\bibinfo {volume} {20}},\ \bibinfo {pages} {1071} (\bibinfo {year}
  {2011}{\natexlab{a}})},\ \Eprint
  {https://arxiv.org/abs/https://doi.org/10.1142/S021830131101840X}
  {https://doi.org/10.1142/S021830131101840X} \BibitemShut {NoStop}%
\bibitem [{\citenamefont {{Somorjai}}\ \emph {et~al.}(1998)\citenamefont
  {{Somorjai}}, \citenamefont {{F{{\"u}}l{{\"o}}p}}, \citenamefont {{Kiss}},
  \citenamefont {{Rolfs}}, \citenamefont {{Trautvetter}}, \citenamefont
  {{Greife}}, \citenamefont {{Junker}}, \citenamefont {{Goriely}},
  \citenamefont {{Arnould}}, \citenamefont {{Rayet}}, \citenamefont
  {{Rauscher}},\ and\ \citenamefont {{Oberhummer}}}]{Somorjai1998}%
  \BibitemOpen
  \bibfield  {author} {\bibinfo {author} {\bibfnamefont {E.}~\bibnamefont
  {{Somorjai}}}, \bibinfo {author} {\bibfnamefont {Z.}~\bibnamefont
  {{F{{\"u}}l{{\"o}}p}}}, \bibinfo {author} {\bibfnamefont {A.~Z.}\
  \bibnamefont {{Kiss}}}, \bibinfo {author} {\bibfnamefont {C.~E.}\
  \bibnamefont {{Rolfs}}}, \bibinfo {author} {\bibfnamefont {H.~P.}\
  \bibnamefont {{Trautvetter}}}, \bibinfo {author} {\bibfnamefont
  {U.}~\bibnamefont {{Greife}}}, \bibinfo {author} {\bibfnamefont
  {M.}~\bibnamefont {{Junker}}}, \bibinfo {author} {\bibfnamefont
  {S.}~\bibnamefont {{Goriely}}}, \bibinfo {author} {\bibfnamefont
  {M.}~\bibnamefont {{Arnould}}}, \bibinfo {author} {\bibfnamefont
  {M.}~\bibnamefont {{Rayet}}}, \bibinfo {author} {\bibfnamefont
  {T.}~\bibnamefont {{Rauscher}}},\ and\ \bibinfo {author} {\bibfnamefont
  {H.}~\bibnamefont {{Oberhummer}}},\ }\href@noop {} {\bibfield  {journal}
  {\bibinfo  {journal} {Astronomy and Astrophysics}\ }\textbf {\bibinfo
  {volume} {333}},\ \bibinfo {pages} {1112} (\bibinfo {year}
  {1998})}\BibitemShut {NoStop}%
\bibitem [{\citenamefont {Scholz}\ \emph {et~al.}(2020)\citenamefont {Scholz},
  \citenamefont {Wilsenach}, \citenamefont {Becker}, \citenamefont {Blazhev},
  \citenamefont {Heim}, \citenamefont {Foteinou}, \citenamefont {Giesen},
  \citenamefont {M{\"u}nker}, \citenamefont {Rogalla}, \citenamefont {Sprung},
  \citenamefont {Zilges},\ and\ \citenamefont {Zuber}}]{Scholz2020}%
  \BibitemOpen
  \bibfield  {author} {\bibinfo {author} {\bibfnamefont {P.}~\bibnamefont
  {Scholz}}, \bibinfo {author} {\bibfnamefont {H.}~\bibnamefont {Wilsenach}},
  \bibinfo {author} {\bibfnamefont {H.~W.}\ \bibnamefont {Becker}}, \bibinfo
  {author} {\bibfnamefont {A.}~\bibnamefont {Blazhev}}, \bibinfo {author}
  {\bibfnamefont {F.}~\bibnamefont {Heim}}, \bibinfo {author} {\bibfnamefont
  {V.}~\bibnamefont {Foteinou}}, \bibinfo {author} {\bibfnamefont
  {U.}~\bibnamefont {Giesen}}, \bibinfo {author} {\bibfnamefont
  {C.}~\bibnamefont {M{\"u}nker}}, \bibinfo {author} {\bibfnamefont
  {D.}~\bibnamefont {Rogalla}}, \bibinfo {author} {\bibfnamefont
  {P.}~\bibnamefont {Sprung}}, \bibinfo {author} {\bibfnamefont
  {A.}~\bibnamefont {Zilges}},\ and\ \bibinfo {author} {\bibfnamefont
  {K.}~\bibnamefont {Zuber}},\ }\href
  {https://doi.org/10.1103/PhysRevC.102.045811} {\bibfield  {journal} {\bibinfo
   {journal} {Phys. Rev. C}\ }\textbf {\bibinfo {volume} {102}},\ \bibinfo
  {pages} {045811} (\bibinfo {year} {2020})}\BibitemShut {NoStop}%
\bibitem [{\citenamefont {Kiss}\ \emph {et~al.}(2022)\citenamefont {Kiss},
  \citenamefont {Mohr}, \citenamefont {Gy\"urky}, \citenamefont {Sz\"ucs},
  \citenamefont {Csedreki}, \citenamefont {Hal\'asz}, \citenamefont
  {F\"ul\"op},\ and\ \citenamefont {Somorjai}}]{Kiss2022}%
  \BibitemOpen
  \bibfield  {author} {\bibinfo {author} {\bibfnamefont {G.~G.}\ \bibnamefont
  {Kiss}}, \bibinfo {author} {\bibfnamefont {P.}~\bibnamefont {Mohr}}, \bibinfo
  {author} {\bibfnamefont {G.}~\bibnamefont {Gy\"urky}}, \bibinfo {author}
  {\bibfnamefont {T.}~\bibnamefont {Sz\"ucs}}, \bibinfo {author} {\bibfnamefont
  {L.}~\bibnamefont {Csedreki}}, \bibinfo {author} {\bibfnamefont
  {Z.}~\bibnamefont {Hal\'asz}}, \bibinfo {author} {\bibfnamefont
  {Z.}~\bibnamefont {F\"ul\"op}},\ and\ \bibinfo {author} {\bibfnamefont
  {E.}~\bibnamefont {Somorjai}},\ }\href
  {https://doi.org/10.1103/PhysRevC.106.015802} {\bibfield  {journal} {\bibinfo
   {journal} {Phys. Rev. C}\ }\textbf {\bibinfo {volume} {106}},\ \bibinfo
  {pages} {015802} (\bibinfo {year} {2022})}\BibitemShut {NoStop}%
\bibitem [{\citenamefont {Denzler}\ \emph {et~al.}(1995)\citenamefont
  {Denzler}, \citenamefont {R{\"o}sch},\ and\ \citenamefont
  {Qaim}}]{Denzler1995}%
  \BibitemOpen
  \bibfield  {author} {\bibinfo {author} {\bibfnamefont {F.-O.}\ \bibnamefont
  {Denzler}}, \bibinfo {author} {\bibfnamefont {F.}~\bibnamefont {R{\"o}sch}},\
  and\ \bibinfo {author} {\bibfnamefont {S.~M.}\ \bibnamefont {Qaim}},\ }\href
  {https://doi.org/doi:10.1524/ract.1995.69.4.209} {\bibfield  {journal}
  {\bibinfo  {journal} {Radiochimica Acta}\ }\textbf {\bibinfo {volume} {69}},\
  \bibinfo {pages} {209} (\bibinfo {year} {1995})}\BibitemShut {NoStop}%
\bibitem [{\citenamefont {Archenti}\ \emph {et~al.}(1989)\citenamefont
  {Archenti}, \citenamefont {Ozafran},\ and\ \citenamefont
  {Nassiff}}]{Archenti1989}%
  \BibitemOpen
  \bibfield  {author} {\bibinfo {author} {\bibfnamefont {A.}~\bibnamefont
  {Archenti}}, \bibinfo {author} {\bibfnamefont {M.~J.}\ \bibnamefont
  {Ozafran}},\ and\ \bibinfo {author} {\bibfnamefont {S.~J.}\ \bibnamefont
  {Nassiff}},\ }\href {https://doi.org/10.1007/BF02060986} {\bibfield
  {journal} {\bibinfo  {journal} {J. Radioanal. Nucl. Chem.}\ }\textbf
  {\bibinfo {volume} {132}},\ \bibinfo {pages} {139} (\bibinfo {year}
  {1989})}\BibitemShut {NoStop}%
\bibitem [{\citenamefont {Nica}(2009)}]{NDS147}%
  \BibitemOpen
  \bibfield  {author} {\bibinfo {author} {\bibfnamefont {N.}~\bibnamefont
  {Nica}},\ }\href {https://doi.org/https://doi.org/10.1016/j.nds.2009.02.003}
  {\bibfield  {journal} {\bibinfo  {journal} {Nuclear Data Sheets}\ }\textbf
  {\bibinfo {volume} {110}},\ \bibinfo {pages} {749} (\bibinfo {year}
  {2009})}\BibitemShut {NoStop}%
\bibitem [{\citenamefont {Gy{\"u}rky}\ \emph {et~al.}(2019)\citenamefont
  {Gy{\"u}rky}, \citenamefont {F{\"u}l{\"o}p}, \citenamefont {K{\"a}ppeler},
  \citenamefont {Kiss},\ and\ \citenamefont {Wallner}}]{Gyurky2019}%
  \BibitemOpen
  \bibfield  {author} {\bibinfo {author} {\bibfnamefont {{\relax
  Gy}.}~\bibnamefont {Gy{\"u}rky}}, \bibinfo {author} {\bibfnamefont {{\relax
  Zs}.}~\bibnamefont {F{\"u}l{\"o}p}}, \bibinfo {author} {\bibfnamefont
  {F.}~\bibnamefont {K{\"a}ppeler}}, \bibinfo {author} {\bibfnamefont {G.~G.}\
  \bibnamefont {Kiss}},\ and\ \bibinfo {author} {\bibfnamefont
  {A.}~\bibnamefont {Wallner}},\ }\href
  {https://doi.org/10.1140/epja/i2019-12708-4} {\bibfield  {journal} {\bibinfo
  {journal} {The European Physical Journal A}\ }\textbf {\bibinfo {volume}
  {55}},\ \bibinfo {pages} {41} (\bibinfo {year} {2019})}\BibitemShut {NoStop}%
\bibitem [{\citenamefont {Rosman}\ and\ \citenamefont
  {Taylor}(1998)}]{Rosman1998}%
  \BibitemOpen
  \bibfield  {author} {\bibinfo {author} {\bibfnamefont {K.~J.~R.}\
  \bibnamefont {Rosman}}\ and\ \bibinfo {author} {\bibfnamefont {P.~D.~P.}\
  \bibnamefont {Taylor}},\ }\href {https://doi.org/doi:10.1351/pac199870010217}
  {\bibfield  {journal} {\bibinfo  {journal} {Pure and Applied Chemistry}\
  }\textbf {\bibinfo {volume} {70}},\ \bibinfo {pages} {217} (\bibinfo {year}
  {1998})}\BibitemShut {NoStop}%
\bibitem [{\citenamefont {Rajta}\ \emph {et~al.}(2018)\citenamefont {Rajta},
  \citenamefont {Vajda}, \citenamefont {Gy{\"u}rky}, \citenamefont {Csedreki},
  \citenamefont {Kiss}, \citenamefont {Biri}, \citenamefont {van Oosterhout},
  \citenamefont {Podaru},\ and\ \citenamefont {Mous}}]{Rajta2018}%
  \BibitemOpen
  \bibfield  {author} {\bibinfo {author} {\bibfnamefont {I.}~\bibnamefont
  {Rajta}}, \bibinfo {author} {\bibfnamefont {I.}~\bibnamefont {Vajda}},
  \bibinfo {author} {\bibfnamefont {{\relax Gy}.}~\bibnamefont {Gy{\"u}rky}},
  \bibinfo {author} {\bibfnamefont {L.}~\bibnamefont {Csedreki}}, \bibinfo
  {author} {\bibfnamefont {A.}~\bibnamefont {Kiss}}, \bibinfo {author}
  {\bibfnamefont {S.}~\bibnamefont {Biri}}, \bibinfo {author} {\bibfnamefont
  {H.}~\bibnamefont {van Oosterhout}}, \bibinfo {author} {\bibfnamefont
  {N.}~\bibnamefont {Podaru}},\ and\ \bibinfo {author} {\bibfnamefont
  {D.}~\bibnamefont {Mous}},\ }\href
  {https://doi.org/https://doi.org/10.1016/j.nima.2017.10.073} {\bibfield
  {journal} {\bibinfo  {journal} {Nuclear Instruments and Methods in Physics
  Research Section A: Accelerators, Spectrometers, Detectors and Associated
  Equipment}\ }\textbf {\bibinfo {volume} {880}},\ \bibinfo {pages} {125 }
  (\bibinfo {year} {2018})}\BibitemShut {NoStop}%
\bibitem [{\citenamefont {Mayer}()}]{SIMNRA}%
  \BibitemOpen
  \bibfield  {author} {\bibinfo {author} {\bibfnamefont {M.}~\bibnamefont
  {Mayer}},\ }\href {https://mam.home.ipp.mpg.de/} {}\bibinfo {note} {\,SIMNRA
  version 6.06}\BibitemShut {NoStop}%
\bibitem [{\citenamefont {{Biri}}\ \emph {et~al.}(2021)\citenamefont {{Biri}},
  \citenamefont {{Vajda}}, \citenamefont {{Hajdu}}, \citenamefont {{R{\'a}cz}},
  \citenamefont {{Cs{\'\i}k}}, \citenamefont {{Korm{\'a}ny}}, \citenamefont
  {{Perduk}}, \citenamefont {{Kocsis}},\ and\ \citenamefont
  {{Rajta}}}]{Biri2021}%
  \BibitemOpen
  \bibfield  {author} {\bibinfo {author} {\bibfnamefont {S.}~\bibnamefont
  {{Biri}}}, \bibinfo {author} {\bibfnamefont {I.~K.}\ \bibnamefont {{Vajda}}},
  \bibinfo {author} {\bibfnamefont {P.}~\bibnamefont {{Hajdu}}}, \bibinfo
  {author} {\bibfnamefont {R.}~\bibnamefont {{R{\'a}cz}}}, \bibinfo {author}
  {\bibfnamefont {A.}~\bibnamefont {{Cs{\'\i}k}}}, \bibinfo {author}
  {\bibfnamefont {Z.}~\bibnamefont {{Korm{\'a}ny}}}, \bibinfo {author}
  {\bibfnamefont {Z.}~\bibnamefont {{Perduk}}}, \bibinfo {author}
  {\bibfnamefont {F.}~\bibnamefont {{Kocsis}}},\ and\ \bibinfo {author}
  {\bibfnamefont {I.}~\bibnamefont {{Rajta}}},\ }\href
  {https://doi.org/10.1140/epjp/s13360-021-01219-z} {\bibfield  {journal}
  {\bibinfo  {journal} {European Physical Journal Plus}\ }\textbf {\bibinfo
  {volume} {136}},\ \bibinfo {eid} {247} (\bibinfo {year} {2021})}\BibitemShut
  {NoStop}%
\bibitem [{\citenamefont {Korkulu}\ \emph {et~al.}(2018)\citenamefont
  {Korkulu}, \citenamefont {\"Ozkan}, \citenamefont {Kiss}, \citenamefont
  {Sz\"ucs}, \citenamefont {Gy\"urky}, \citenamefont {F\"ul\"op}, \citenamefont
  {G\"uray}, \citenamefont {Hal\'asz}, \citenamefont {Rauscher}, \citenamefont
  {Somorjai}, \citenamefont {T\"or\"ok},\ and\ \citenamefont {Yal\ifmmode
  \mbox{\c{c}}\else \c{c}\fi{}\ifmmode \imath \else~\i \fi{}n}}]{Korkulu2018}%
  \BibitemOpen
  \bibfield  {author} {\bibinfo {author} {\bibfnamefont {Z.}~\bibnamefont
  {Korkulu}}, \bibinfo {author} {\bibfnamefont {N.}~\bibnamefont {\"Ozkan}},
  \bibinfo {author} {\bibfnamefont {G.~G.}\ \bibnamefont {Kiss}}, \bibinfo
  {author} {\bibfnamefont {T.}~\bibnamefont {Sz\"ucs}}, \bibinfo {author}
  {\bibfnamefont {G.}~\bibnamefont {Gy\"urky}}, \bibinfo {author}
  {\bibfnamefont {Z.}~\bibnamefont {F\"ul\"op}}, \bibinfo {author}
  {\bibfnamefont {R.~T.}\ \bibnamefont {G\"uray}}, \bibinfo {author}
  {\bibfnamefont {Z.}~\bibnamefont {Hal\'asz}}, \bibinfo {author}
  {\bibfnamefont {T.}~\bibnamefont {Rauscher}}, \bibinfo {author}
  {\bibfnamefont {E.}~\bibnamefont {Somorjai}}, \bibinfo {author}
  {\bibfnamefont {Z.}~\bibnamefont {T\"or\"ok}},\ and\ \bibinfo {author}
  {\bibfnamefont {C.}~\bibnamefont {Yal\ifmmode \mbox{\c{c}}\else
  \c{c}\fi{}\ifmmode \imath \else~\i \fi{}n}},\ }\href
  {https://doi.org/10.1103/PhysRevC.97.045803} {\bibfield  {journal} {\bibinfo
  {journal} {Phys. Rev. C}\ }\textbf {\bibinfo {volume} {97}},\ \bibinfo
  {pages} {045803} (\bibinfo {year} {2018})}\BibitemShut {NoStop}%
\bibitem [{\citenamefont {Gy{\"u}rky}\ \emph {et~al.}(2021)\citenamefont
  {Gy{\"u}rky}, \citenamefont {Hal{\'{a}}sz}, \citenamefont {Kiss},
  \citenamefont {Sz{\"u}cs}, \citenamefont {Husz{\'{a}}nk}, \citenamefont
  {T{\"o}r{\"o}k}, \citenamefont {F{\"u}l{\"o}p}, \citenamefont {Rauscher},\
  and\ \citenamefont {Travaglio}}]{Gyurky2021}%
  \BibitemOpen
  \bibfield  {author} {\bibinfo {author} {\bibfnamefont {G.}~\bibnamefont
  {Gy{\"u}rky}}, \bibinfo {author} {\bibfnamefont {Z.}~\bibnamefont
  {Hal{\'{a}}sz}}, \bibinfo {author} {\bibfnamefont {G.~G.}\ \bibnamefont
  {Kiss}}, \bibinfo {author} {\bibfnamefont {T.}~\bibnamefont {Sz{\"u}cs}},
  \bibinfo {author} {\bibfnamefont {R.}~\bibnamefont {Husz{\'{a}}nk}}, \bibinfo
  {author} {\bibfnamefont {Z.}~\bibnamefont {T{\"o}r{\"o}k}}, \bibinfo {author}
  {\bibfnamefont {Z.}~\bibnamefont {F{\"u}l{\"o}p}}, \bibinfo {author}
  {\bibfnamefont {T.}~\bibnamefont {Rauscher}},\ and\ \bibinfo {author}
  {\bibfnamefont {C.}~\bibnamefont {Travaglio}},\ }\href
  {https://doi.org/10.1088/1361-6471/ac2132} {\bibfield  {journal} {\bibinfo
  {journal} {Journal of Physics G: Nuclear and Particle Physics}\ }\textbf
  {\bibinfo {volume} {48}},\ \bibinfo {pages} {105202} (\bibinfo {year}
  {2021})}\BibitemShut {NoStop}%
\bibitem [{\citenamefont {Sz{\"u}cs}\ \emph {et~al.}(2016)\citenamefont
  {Sz{\"u}cs}, \citenamefont {Kiss}, \citenamefont {Rauscher}, \citenamefont
  {T{\"o}{\"o}k}, \citenamefont {Hal{\'{a}}sz}, \citenamefont {F{\"u}l{\"o}p},
  \citenamefont {Gy{\"u}rky},\ and\ \citenamefont {Somorjai}}]{Szucs2016}%
  \BibitemOpen
  \bibfield  {author} {\bibinfo {author} {\bibfnamefont {T.}~\bibnamefont
  {Sz{\"u}cs}}, \bibinfo {author} {\bibfnamefont {G.}~\bibnamefont {Kiss}},
  \bibinfo {author} {\bibfnamefont {T.}~\bibnamefont {Rauscher}}, \bibinfo
  {author} {\bibfnamefont {Z.}~\bibnamefont {T{\"o}{\"o}k}}, \bibinfo {author}
  {\bibfnamefont {Z.}~\bibnamefont {Hal{\'{a}}sz}}, \bibinfo {author}
  {\bibfnamefont {Z.}~\bibnamefont {F{\"u}l{\"o}p}}, \bibinfo {author}
  {\bibfnamefont {G.}~\bibnamefont {Gy{\"u}rky}},\ and\ \bibinfo {author}
  {\bibfnamefont {E.}~\bibnamefont {Somorjai}},\ }\href
  {https://doi.org/10.1088/1742-6596/665/1/012041} {\bibfield  {journal}
  {\bibinfo  {journal} {Journal of Physics: Conference Series}\ }\textbf
  {\bibinfo {volume} {665}},\ \bibinfo {pages} {012041} (\bibinfo {year}
  {2016})}\BibitemShut {NoStop}%
\bibitem [{\citenamefont {Mohr}\ \emph {et~al.}(1997)\citenamefont {Mohr},
  \citenamefont {Rauscher}, \citenamefont {Oberhummer}, \citenamefont
  {M\'at\'e}, \citenamefont {F\"ul\"op}, \citenamefont {Somorjai},
  \citenamefont {Jaeger},\ and\ \citenamefont {Staudt}}]{Mohr1997}%
  \BibitemOpen
  \bibfield  {author} {\bibinfo {author} {\bibfnamefont {P.}~\bibnamefont
  {Mohr}}, \bibinfo {author} {\bibfnamefont {T.}~\bibnamefont {Rauscher}},
  \bibinfo {author} {\bibfnamefont {H.}~\bibnamefont {Oberhummer}}, \bibinfo
  {author} {\bibfnamefont {Z.}~\bibnamefont {M\'at\'e}}, \bibinfo {author}
  {\bibfnamefont {Z.}~\bibnamefont {F\"ul\"op}}, \bibinfo {author}
  {\bibfnamefont {E.}~\bibnamefont {Somorjai}}, \bibinfo {author}
  {\bibfnamefont {M.}~\bibnamefont {Jaeger}},\ and\ \bibinfo {author}
  {\bibfnamefont {G.}~\bibnamefont {Staudt}},\ }\href
  {https://doi.org/10.1103/PhysRevC.55.1523} {\bibfield  {journal} {\bibinfo
  {journal} {Phys. Rev. C}\ }\textbf {\bibinfo {volume} {55}},\ \bibinfo
  {pages} {1523} (\bibinfo {year} {1997})}\BibitemShut {NoStop}%
\bibitem [{\citenamefont {Koning}\ \emph {et~al.}()\citenamefont {Koning},
  \citenamefont {Hilaire},\ and\ \citenamefont {Goriely}}]{TALYS}%
  \BibitemOpen
  \bibfield  {author} {\bibinfo {author} {\bibfnamefont {A.~J.}\ \bibnamefont
  {Koning}}, \bibinfo {author} {\bibfnamefont {S.}~\bibnamefont {Hilaire}},\
  and\ \bibinfo {author} {\bibfnamefont {S.}~\bibnamefont {Goriely}},\ }\href
  {http://www.talys.eu/} {\bibinfo {title} {computer code {\sc{talys}}, version
  1.9}}\BibitemShut {NoStop}%
\bibitem [{\citenamefont {Rauscher}(2011{\natexlab{b}})}]{Rauscher_IJMPE2011}%
  \BibitemOpen
  \bibfield  {author} {\bibinfo {author} {\bibfnamefont {T.}~\bibnamefont
  {Rauscher}},\ }\href {https://doi.org/10.1142/S021830131101840X} {\bibfield
  {journal} {\bibinfo  {journal} {International Journal of Modern Physics E}\
  }\textbf {\bibinfo {volume} {20}},\ \bibinfo {pages} {1071} (\bibinfo {year}
  {2011}{\natexlab{b}})}\BibitemShut {NoStop}%
\bibitem [{\citenamefont {Mohr}\ \emph {et~al.}(2020)\citenamefont {Mohr},
  \citenamefont {F\"ul\"op}, \citenamefont {Gy\"urky}, \citenamefont {Kiss},\
  and\ \citenamefont {Sz\"ucs}}]{Mohr_PRL2020}%
  \BibitemOpen
  \bibfield  {author} {\bibinfo {author} {\bibfnamefont {P.}~\bibnamefont
  {Mohr}}, \bibinfo {author} {\bibfnamefont {Z.}~\bibnamefont {F\"ul\"op}},
  \bibinfo {author} {\bibfnamefont {G.}~\bibnamefont {Gy\"urky}}, \bibinfo
  {author} {\bibfnamefont {G.~G.}\ \bibnamefont {Kiss}},\ and\ \bibinfo
  {author} {\bibfnamefont {T.}~\bibnamefont {Sz\"ucs}},\ }\href
  {https://doi.org/10.1103/PhysRevLett.124.252701} {\bibfield  {journal}
  {\bibinfo  {journal} {Phys. Rev. Lett.}\ }\textbf {\bibinfo {volume} {124}},\
  \bibinfo {pages} {252701} (\bibinfo {year} {2020})}\BibitemShut {NoStop}%
\bibitem [{\citenamefont {Mohr}\ \emph {et~al.}(2021)\citenamefont {Mohr},
  \citenamefont {F{\"u}l{\"o}p}, \citenamefont {Gy{\"u}rky}, \citenamefont
  {Kiss}, \citenamefont {Sz{\"u}cs}, \citenamefont {Arcones}, \citenamefont
  {Jacobi},\ and\ \citenamefont {Psaltis}}]{Mohr_ADNDT2021}%
  \BibitemOpen
  \bibfield  {author} {\bibinfo {author} {\bibfnamefont {P.}~\bibnamefont
  {Mohr}}, \bibinfo {author} {\bibfnamefont {Z.}~\bibnamefont {F{\"u}l{\"o}p}},
  \bibinfo {author} {\bibfnamefont {G.}~\bibnamefont {Gy{\"u}rky}}, \bibinfo
  {author} {\bibfnamefont {G.}~\bibnamefont {Kiss}}, \bibinfo {author}
  {\bibfnamefont {T.}~\bibnamefont {Sz{\"u}cs}}, \bibinfo {author}
  {\bibfnamefont {A.}~\bibnamefont {Arcones}}, \bibinfo {author} {\bibfnamefont
  {M.}~\bibnamefont {Jacobi}},\ and\ \bibinfo {author} {\bibfnamefont
  {A.}~\bibnamefont {Psaltis}},\ }\href
  {https://doi.org/https://doi.org/10.1016/j.adt.2021.101453} {\bibfield
  {journal} {\bibinfo  {journal} {Atomic Data and Nuclear Data Tables}\
  }\textbf {\bibinfo {volume} {142}},\ \bibinfo {pages} {101453} (\bibinfo
  {year} {2021})}\BibitemShut {NoStop}%
\bibitem [{\citenamefont {Avrigeanu}\ \emph {et~al.}(2014)\citenamefont
  {Avrigeanu}, \citenamefont {Avrigeanu},\ and\ \citenamefont {M\ifmmode
  \u{a}\else \u{a}\fi{}n\ifmmode~\u{a}\else
  \u{a}\fi{}ilescu}}]{Avrigeanu_PRC2014_aomp}%
  \BibitemOpen
  \bibfield  {author} {\bibinfo {author} {\bibfnamefont {V.}~\bibnamefont
  {Avrigeanu}}, \bibinfo {author} {\bibfnamefont {M.}~\bibnamefont
  {Avrigeanu}},\ and\ \bibinfo {author} {\bibfnamefont {C.}~\bibnamefont
  {M\ifmmode \u{a}\else \u{a}\fi{}n\ifmmode~\u{a}\else \u{a}\fi{}ilescu}},\
  }\href {https://doi.org/10.1103/PhysRevC.90.044612} {\bibfield  {journal}
  {\bibinfo  {journal} {Phys. Rev. C}\ }\textbf {\bibinfo {volume} {90}},\
  \bibinfo {pages} {044612} (\bibinfo {year} {2014})}\BibitemShut {NoStop}%
\bibitem [{\citenamefont {McFadden}\ and\ \citenamefont
  {Satchler}(1966)}]{McFadden_NPA1966_aomp}%
  \BibitemOpen
  \bibfield  {author} {\bibinfo {author} {\bibfnamefont {L.}~\bibnamefont
  {McFadden}}\ and\ \bibinfo {author} {\bibfnamefont {G.~R.}\ \bibnamefont
  {Satchler}},\ }\href
  {https://doi.org/https://doi.org/10.1016/0029-5582(66)90441-X} {\bibfield
  {journal} {\bibinfo  {journal} {Nuclear Physics}\ }\textbf {\bibinfo {volume}
  {84}},\ \bibinfo {pages} {177} (\bibinfo {year} {1966})}\BibitemShut
  {NoStop}%
\bibitem [{\citenamefont {Demetriou}\ \emph {et~al.}(2002)\citenamefont
  {Demetriou}, \citenamefont {Grama},\ and\ \citenamefont
  {Goriely}}]{Demetriou_NPA2002_aomp}%
  \BibitemOpen
  \bibfield  {author} {\bibinfo {author} {\bibfnamefont {P.}~\bibnamefont
  {Demetriou}}, \bibinfo {author} {\bibfnamefont {C.}~\bibnamefont {Grama}},\
  and\ \bibinfo {author} {\bibfnamefont {S.}~\bibnamefont {Goriely}},\ }\href
  {https://doi.org/https://doi.org/10.1016/S0375-9474(02)00756-X} {\bibfield
  {journal} {\bibinfo  {journal} {Nuclear Physics A}\ }\textbf {\bibinfo
  {volume} {707}},\ \bibinfo {pages} {253} (\bibinfo {year}
  {2002})}\BibitemShut {NoStop}%
\bibitem [{\citenamefont {Sz\"ucs}\ \emph {et~al.}(2019)\citenamefont
  {Sz\"ucs}, \citenamefont {Mohr}, \citenamefont {Gy\"urky}, \citenamefont
  {Hal\'asz}, \citenamefont {Husz\'ank}, \citenamefont {Kiss}, \citenamefont
  {Szegedi}, \citenamefont {T\"or\"ok},\ and\ \citenamefont
  {F\"ul\"op}}]{Szucs_PRC2019_au197aX}%
  \BibitemOpen
  \bibfield  {author} {\bibinfo {author} {\bibfnamefont {T.}~\bibnamefont
  {Sz\"ucs}}, \bibinfo {author} {\bibfnamefont {P.}~\bibnamefont {Mohr}},
  \bibinfo {author} {\bibfnamefont {G.}~\bibnamefont {Gy\"urky}}, \bibinfo
  {author} {\bibfnamefont {Z.}~\bibnamefont {Hal\'asz}}, \bibinfo {author}
  {\bibfnamefont {R.}~\bibnamefont {Husz\'ank}}, \bibinfo {author}
  {\bibfnamefont {G.~G.}\ \bibnamefont {Kiss}}, \bibinfo {author}
  {\bibfnamefont {T.~N.}\ \bibnamefont {Szegedi}}, \bibinfo {author}
  {\bibfnamefont {Z.}~\bibnamefont {T\"or\"ok}},\ and\ \bibinfo {author}
  {\bibfnamefont {Z.}~\bibnamefont {F\"ul\"op}},\ }\href
  {https://doi.org/10.1103/PhysRevC.100.065803} {\bibfield  {journal} {\bibinfo
   {journal} {Phys. Rev. C}\ }\textbf {\bibinfo {volume} {100}},\ \bibinfo
  {pages} {065803} (\bibinfo {year} {2019})}\BibitemShut {NoStop}%
\bibitem [{\citenamefont {Mohr}\ \emph {et~al.}(2018)\citenamefont {Mohr},
  \citenamefont {Talwar},\ and\ \citenamefont {Avila}}]{Mohr_PRC2018_38ar}%
  \BibitemOpen
  \bibfield  {author} {\bibinfo {author} {\bibfnamefont {P.}~\bibnamefont
  {Mohr}}, \bibinfo {author} {\bibfnamefont {R.}~\bibnamefont {Talwar}},\ and\
  \bibinfo {author} {\bibfnamefont {M.~L.}\ \bibnamefont {Avila}},\ }\href
  {https://doi.org/10.1103/PhysRevC.98.045805} {\bibfield  {journal} {\bibinfo
  {journal} {Phys. Rev. C}\ }\textbf {\bibinfo {volume} {98}},\ \bibinfo
  {pages} {045805} (\bibinfo {year} {2018})}\BibitemShut {NoStop}%
\bibitem [{\citenamefont {Mohr}\ \emph {et~al.}(2017)\citenamefont {Mohr},
  \citenamefont {Gy\"urky},\ and\ \citenamefont
  {F\"ul\"op}}]{Mohr_PRC2017_64zn}%
  \BibitemOpen
  \bibfield  {author} {\bibinfo {author} {\bibfnamefont {P.}~\bibnamefont
  {Mohr}}, \bibinfo {author} {\bibfnamefont {G.}~\bibnamefont {Gy\"urky}},\
  and\ \bibinfo {author} {\bibfnamefont {Z.}~\bibnamefont {F\"ul\"op}},\ }\href
  {https://doi.org/10.1103/PhysRevC.95.015807} {\bibfield  {journal} {\bibinfo
  {journal} {Phys. Rev. C}\ }\textbf {\bibinfo {volume} {95}},\ \bibinfo
  {pages} {015807} (\bibinfo {year} {2017})}\BibitemShut {NoStop}%
\bibitem [{\citenamefont {Watanabe}(1958)}]{Watanabe_NPA1958_omp}%
  \BibitemOpen
  \bibfield  {author} {\bibinfo {author} {\bibfnamefont {S.}~\bibnamefont
  {Watanabe}},\ }\href
  {https://doi.org/https://doi.org/10.1016/0029-5582(58)90180-9} {\bibfield
  {journal} {\bibinfo  {journal} {Nuclear Physics}\ }\textbf {\bibinfo {volume}
  {8}},\ \bibinfo {pages} {484} (\bibinfo {year} {1958})}\BibitemShut {NoStop}%
\bibitem [{\citenamefont {Nolte}\ \emph {et~al.}(1987)\citenamefont {Nolte},
  \citenamefont {Machner},\ and\ \citenamefont
  {Bojowald}}]{Nolte_PRC1987_aomp}%
  \BibitemOpen
  \bibfield  {author} {\bibinfo {author} {\bibfnamefont {M.}~\bibnamefont
  {Nolte}}, \bibinfo {author} {\bibfnamefont {H.}~\bibnamefont {Machner}},\
  and\ \bibinfo {author} {\bibfnamefont {J.}~\bibnamefont {Bojowald}},\ }\href
  {https://doi.org/10.1103/PhysRevC.36.1312} {\bibfield  {journal} {\bibinfo
  {journal} {Phys. Rev. C}\ }\textbf {\bibinfo {volume} {36}},\ \bibinfo
  {pages} {1312} (\bibinfo {year} {1987})}\BibitemShut {NoStop}%
\bibitem [{\citenamefont {Avrigeanu}\ \emph {et~al.}(1994)\citenamefont
  {Avrigeanu}, \citenamefont {Hodgson},\ and\ \citenamefont
  {Avrigeanu}}]{Avrigeanu_PRC1994_aomp}%
  \BibitemOpen
  \bibfield  {author} {\bibinfo {author} {\bibfnamefont {V.}~\bibnamefont
  {Avrigeanu}}, \bibinfo {author} {\bibfnamefont {P.~E.}\ \bibnamefont
  {Hodgson}},\ and\ \bibinfo {author} {\bibfnamefont {M.}~\bibnamefont
  {Avrigeanu}},\ }\href {https://doi.org/10.1103/PhysRevC.49.2136} {\bibfield
  {journal} {\bibinfo  {journal} {Phys. Rev. C}\ }\textbf {\bibinfo {volume}
  {49}},\ \bibinfo {pages} {2136} (\bibinfo {year} {1994})}\BibitemShut
  {NoStop}%
\bibitem [{\citenamefont {Rauscher}\ and\ \citenamefont
  {Thielemann}(2000)}]{Rauscher_ADNDT2000_rates}%
  \BibitemOpen
  \bibfield  {author} {\bibinfo {author} {\bibfnamefont {T.}~\bibnamefont
  {Rauscher}}\ and\ \bibinfo {author} {\bibfnamefont {F.-K.}\ \bibnamefont
  {Thielemann}},\ }\href
  {https://doi.org/https://doi.org/10.1006/adnd.2000.0834} {\bibfield
  {journal} {\bibinfo  {journal} {Atomic Data and Nuclear Data Tables}\
  }\textbf {\bibinfo {volume} {75}},\ \bibinfo {pages} {1} (\bibinfo {year}
  {2000})}\BibitemShut {NoStop}%
\bibitem [{\citenamefont {Cyburt}\ \emph {et~al.}(2010)\citenamefont {Cyburt},
  \citenamefont {Amthor}, \citenamefont {Ferguson}, \citenamefont {Meisel},
  \citenamefont {Smith}, \citenamefont {Warren}, \citenamefont {Heger},
  \citenamefont {Hoffman}, \citenamefont {Rauscher}, \citenamefont {Sakharuk},
  \citenamefont {Schatz}, \citenamefont {Thielemann},\ and\ \citenamefont
  {Wiescher}}]{Cyburt_APJS2010_reaclib}%
  \BibitemOpen
  \bibfield  {author} {\bibinfo {author} {\bibfnamefont {R.~H.}\ \bibnamefont
  {Cyburt}}, \bibinfo {author} {\bibfnamefont {A.~M.}\ \bibnamefont {Amthor}},
  \bibinfo {author} {\bibfnamefont {R.}~\bibnamefont {Ferguson}}, \bibinfo
  {author} {\bibfnamefont {Z.}~\bibnamefont {Meisel}}, \bibinfo {author}
  {\bibfnamefont {K.}~\bibnamefont {Smith}}, \bibinfo {author} {\bibfnamefont
  {S.}~\bibnamefont {Warren}}, \bibinfo {author} {\bibfnamefont
  {A.}~\bibnamefont {Heger}}, \bibinfo {author} {\bibfnamefont {R.~D.}\
  \bibnamefont {Hoffman}}, \bibinfo {author} {\bibfnamefont {T.}~\bibnamefont
  {Rauscher}}, \bibinfo {author} {\bibfnamefont {A.}~\bibnamefont {Sakharuk}},
  \bibinfo {author} {\bibfnamefont {H.}~\bibnamefont {Schatz}}, \bibinfo
  {author} {\bibfnamefont {F.~K.}\ \bibnamefont {Thielemann}},\ and\ \bibinfo
  {author} {\bibfnamefont {M.}~\bibnamefont {Wiescher}},\ }\href
  {http://stacks.iop.org/0067-0049/189/i=1/a=240} {\bibfield  {journal}
  {\bibinfo  {journal} {The Astrophysical Journal Supplement Series}\ }\textbf
  {\bibinfo {volume} {189}},\ \bibinfo {pages} {240} (\bibinfo {year}
  {2010})}\BibitemShut {NoStop}%
\bibitem [{\citenamefont {Mohr}(2015)}]{Mohr_EPJA2015_A20-50}%
  \BibitemOpen
  \bibfield  {author} {\bibinfo {author} {\bibfnamefont {P.}~\bibnamefont
  {Mohr}},\ }\href {https://doi.org/10.1140/epja/i2015-15056-5} {\bibfield
  {journal} {\bibinfo  {journal} {The European Physical Journal A}\ }\textbf
  {\bibinfo {volume} {51}},\ \bibinfo {pages} {56} (\bibinfo {year}
  {2015})}\BibitemShut {NoStop}%
\bibitem [{\citenamefont {Sauerwein}\ \emph {et~al.}(2011)\citenamefont
  {Sauerwein}, \citenamefont {Becker}, \citenamefont {Dombrowski},
  \citenamefont {Elvers}, \citenamefont {Endres}, \citenamefont {Giesen},
  \citenamefont {Hasper}, \citenamefont {Hennig}, \citenamefont {Netterdon},
  \citenamefont {Rauscher}, \citenamefont {Rogalla}, \citenamefont {Zell},\
  and\ \citenamefont {Zilges}}]{Sauerwein_PRC2011_141pr}%
  \BibitemOpen
  \bibfield  {author} {\bibinfo {author} {\bibfnamefont {A.}~\bibnamefont
  {Sauerwein}}, \bibinfo {author} {\bibfnamefont {H.-W.}\ \bibnamefont
  {Becker}}, \bibinfo {author} {\bibfnamefont {H.}~\bibnamefont {Dombrowski}},
  \bibinfo {author} {\bibfnamefont {M.}~\bibnamefont {Elvers}}, \bibinfo
  {author} {\bibfnamefont {J.}~\bibnamefont {Endres}}, \bibinfo {author}
  {\bibfnamefont {U.}~\bibnamefont {Giesen}}, \bibinfo {author} {\bibfnamefont
  {J.}~\bibnamefont {Hasper}}, \bibinfo {author} {\bibfnamefont
  {A.}~\bibnamefont {Hennig}}, \bibinfo {author} {\bibfnamefont
  {L.}~\bibnamefont {Netterdon}}, \bibinfo {author} {\bibfnamefont
  {T.}~\bibnamefont {Rauscher}}, \bibinfo {author} {\bibfnamefont
  {D.}~\bibnamefont {Rogalla}}, \bibinfo {author} {\bibfnamefont {K.~O.}\
  \bibnamefont {Zell}},\ and\ \bibinfo {author} {\bibfnamefont
  {A.}~\bibnamefont {Zilges}},\ }\href
  {https://doi.org/10.1103/PhysRevC.84.045808} {\bibfield  {journal} {\bibinfo
  {journal} {Phys. Rev. C}\ }\textbf {\bibinfo {volume} {84}},\ \bibinfo
  {pages} {045808} (\bibinfo {year} {2011})}\BibitemShut {NoStop}%
\bibitem [{\citenamefont {Mohr}(2011)}]{Mohr_PRC2011_pr141a}%
  \BibitemOpen
  \bibfield  {author} {\bibinfo {author} {\bibfnamefont {P.}~\bibnamefont
  {Mohr}},\ }\href {https://doi.org/10.1103/PhysRevC.84.055803} {\bibfield
  {journal} {\bibinfo  {journal} {Phys. Rev. C}\ }\textbf {\bibinfo {volume}
  {84}},\ \bibinfo {pages} {055803} (\bibinfo {year} {2011})}\BibitemShut
  {NoStop}%
\bibitem [{\citenamefont {Sallaska}\ \emph {et~al.}(2013)\citenamefont
  {Sallaska}, \citenamefont {Iliadis}, \citenamefont {Champagne}, \citenamefont
  {Goriely}, \citenamefont {Starrfield},\ and\ \citenamefont
  {Timmes}}]{Sallaska_APJS2013_starlib}%
  \BibitemOpen
  \bibfield  {author} {\bibinfo {author} {\bibfnamefont {A.~L.}\ \bibnamefont
  {Sallaska}}, \bibinfo {author} {\bibfnamefont {C.}~\bibnamefont {Iliadis}},
  \bibinfo {author} {\bibfnamefont {A.~E.}\ \bibnamefont {Champagne}}, \bibinfo
  {author} {\bibfnamefont {S.}~\bibnamefont {Goriely}}, \bibinfo {author}
  {\bibfnamefont {S.}~\bibnamefont {Starrfield}},\ and\ \bibinfo {author}
  {\bibfnamefont {F.~X.}\ \bibnamefont {Timmes}},\ }\href
  {http://stacks.iop.org/0067-0049/207/i=1/a=18} {\bibfield  {journal}
  {\bibinfo  {journal} {The Astrophysical Journal Supplement Series}\ }\textbf
  {\bibinfo {volume} {207}},\ \bibinfo {pages} {18} (\bibinfo {year}
  {2013})}\BibitemShut {NoStop}%
\bibitem [{\citenamefont {Avrigeanu}\ \emph {et~al.}(2003)\citenamefont
  {Avrigeanu}, \citenamefont {von Oertzen}, \citenamefont {Plompen},\ and\
  \citenamefont {Avrigeanu}}]{Avrigeanu_NPA2003_aomp}%
  \BibitemOpen
  \bibfield  {author} {\bibinfo {author} {\bibfnamefont {M.}~\bibnamefont
  {Avrigeanu}}, \bibinfo {author} {\bibfnamefont {W.}~\bibnamefont
  {von Oertzen}}, \bibinfo {author} {\bibfnamefont {A.}~\bibnamefont
  {Plompen}},\ and\ \bibinfo {author} {\bibfnamefont {V.}~\bibnamefont
  {Avrigeanu}},\ }\href
  {https://doi.org/https://doi.org/10.1016/S0375-9474(03)01159-X} {\bibfield
  {journal} {\bibinfo  {journal} {Nuclear Physics A}\ }\textbf {\bibinfo
  {volume} {723}},\ \bibinfo {pages} {104} (\bibinfo {year}
  {2003})}\BibitemShut {NoStop}%
\bibitem [{\citenamefont {Mohr}\ \emph {et~al.}(2013)\citenamefont {Mohr},
  \citenamefont {Kiss}, \citenamefont {F{\"u}l{\"o}p}, \citenamefont {Galaviz},
  \citenamefont {Gy{\"u}rky},\ and\ \citenamefont
  {Somorjai}}]{Mohr_ADNDT2013_atomki-v1}%
  \BibitemOpen
  \bibfield  {author} {\bibinfo {author} {\bibfnamefont {P.}~\bibnamefont
  {Mohr}}, \bibinfo {author} {\bibfnamefont {G.}~\bibnamefont {Kiss}}, \bibinfo
  {author} {\bibfnamefont {Z.}~\bibnamefont {F{\"u}l{\"o}p}}, \bibinfo {author}
  {\bibfnamefont {D.}~\bibnamefont {Galaviz}}, \bibinfo {author} {\bibfnamefont
  {G.}~\bibnamefont {Gy{\"u}rky}},\ and\ \bibinfo {author} {\bibfnamefont
  {E.}~\bibnamefont {Somorjai}},\ }\href
  {https://doi.org/https://doi.org/10.1016/j.adt.2012.10.003} {\bibfield
  {journal} {\bibinfo  {journal} {Atomic Data and Nuclear Data Tables}\
  }\textbf {\bibinfo {volume} {99}},\ \bibinfo {pages} {651} (\bibinfo {year}
  {2013})}\BibitemShut {NoStop}%
\bibitem [{\citenamefont {Fang}\ \emph {et~al.}(2022)\citenamefont {Fang},
  \citenamefont {Frossard}, \citenamefont {Boyet}, \citenamefont {Bouvier},
  \citenamefont {Barrat}, \citenamefont {Chaussidon},\ and\ \citenamefont
  {Moynier}}]{LinruFang2022}%
  \BibitemOpen
  \bibfield  {author} {\bibinfo {author} {\bibfnamefont {L.}~\bibnamefont
  {Fang}}, \bibinfo {author} {\bibfnamefont {P.}~\bibnamefont {Frossard}},
  \bibinfo {author} {\bibfnamefont {M.}~\bibnamefont {Boyet}}, \bibinfo
  {author} {\bibfnamefont {A.}~\bibnamefont {Bouvier}}, \bibinfo {author}
  {\bibfnamefont {J.-A.}\ \bibnamefont {Barrat}}, \bibinfo {author}
  {\bibfnamefont {M.}~\bibnamefont {Chaussidon}},\ and\ \bibinfo {author}
  {\bibfnamefont {F.}~\bibnamefont {Moynier}},\ }\href
  {https://doi.org/10.1073/pnas.2120933119} {\bibfield  {journal} {\bibinfo
  {journal} {Proceedings of the National Academy of Sciences}\ }\textbf
  {\bibinfo {volume} {119}},\ \bibinfo {pages} {e2120933119} (\bibinfo {year}
  {2022})},\ \Eprint
  {https://arxiv.org/abs/https://www.pnas.org/doi/pdf/10.1073/pnas.2120933119}
  {https://www.pnas.org/doi/pdf/10.1073/pnas.2120933119} \BibitemShut {NoStop}%
\bibitem [{\citenamefont {Lugaro}\ \emph {et~al.}(2016)\citenamefont {Lugaro},
  \citenamefont {Pignatari}, \citenamefont {Ott}, \citenamefont {Zuber},
  \citenamefont {Travaglio}, \citenamefont {Gy{\"u}rky},\ and\ \citenamefont
  {F{\"u}l{\"o}p}}]{Lugaro2016_PNAS}%
  \BibitemOpen
  \bibfield  {author} {\bibinfo {author} {\bibfnamefont {M.}~\bibnamefont
  {Lugaro}}, \bibinfo {author} {\bibfnamefont {M.}~\bibnamefont {Pignatari}},
  \bibinfo {author} {\bibfnamefont {U.}~\bibnamefont {Ott}}, \bibinfo {author}
  {\bibfnamefont {K.}~\bibnamefont {Zuber}}, \bibinfo {author} {\bibfnamefont
  {C.}~\bibnamefont {Travaglio}}, \bibinfo {author} {\bibfnamefont
  {G.}~\bibnamefont {Gy{\"u}rky}},\ and\ \bibinfo {author} {\bibfnamefont
  {Z.}~\bibnamefont {F{\"u}l{\"o}p}},\ }\href
  {https://doi.org/10.1073/pnas.1519344113} {\bibfield  {journal} {\bibinfo
  {journal} {Proceedings of the National Academy of Sciences}\ }\textbf
  {\bibinfo {volume} {113}},\ \bibinfo {pages} {907} (\bibinfo {year}
  {2016})},\ \Eprint
  {https://arxiv.org/abs/https://www.pnas.org/doi/pdf/10.1073/pnas.1519344113}
  {https://www.pnas.org/doi/pdf/10.1073/pnas.1519344113} \BibitemShut {NoStop}%
\bibitem [{\citenamefont {Rauscher}(2013)}]{Rauscher2013_PRL}%
  \BibitemOpen
  \bibfield  {author} {\bibinfo {author} {\bibfnamefont {T.}~\bibnamefont
  {Rauscher}},\ }\href {https://doi.org/10.1103/PhysRevLett.111.061104}
  {\bibfield  {journal} {\bibinfo  {journal} {Phys. Rev. Lett.}\ }\textbf
  {\bibinfo {volume} {111}},\ \bibinfo {pages} {061104} (\bibinfo {year}
  {2013})}\BibitemShut {NoStop}%
\bibitem [{\citenamefont {Kinoshita}\ \emph {et~al.}(2012)\citenamefont
  {Kinoshita}, \citenamefont {Paul}, \citenamefont {Kashiv}, \citenamefont
  {Collon}, \citenamefont {Deibel}, \citenamefont {DiGiovine}, \citenamefont
  {Greene}, \citenamefont {Henderson}, \citenamefont {Jiang}, \citenamefont
  {Marley}, \citenamefont {Nakanishi}, \citenamefont {Pardo}, \citenamefont
  {Rehm}, \citenamefont {Robertson}, \citenamefont {Scott}, \citenamefont
  {Schmitt}, \citenamefont {Tang}, \citenamefont {Vondrasek},\ and\
  \citenamefont {Yokoyama}}]{Kinoshita2012_Science}%
  \BibitemOpen
  \bibfield  {author} {\bibinfo {author} {\bibfnamefont {N.}~\bibnamefont
  {Kinoshita}}, \bibinfo {author} {\bibfnamefont {M.}~\bibnamefont {Paul}},
  \bibinfo {author} {\bibfnamefont {Y.}~\bibnamefont {Kashiv}}, \bibinfo
  {author} {\bibfnamefont {P.}~\bibnamefont {Collon}}, \bibinfo {author}
  {\bibfnamefont {C.~M.}\ \bibnamefont {Deibel}}, \bibinfo {author}
  {\bibfnamefont {B.}~\bibnamefont {DiGiovine}}, \bibinfo {author}
  {\bibfnamefont {J.~P.}\ \bibnamefont {Greene}}, \bibinfo {author}
  {\bibfnamefont {D.~J.}\ \bibnamefont {Henderson}}, \bibinfo {author}
  {\bibfnamefont {C.~L.}\ \bibnamefont {Jiang}}, \bibinfo {author}
  {\bibfnamefont {S.~T.}\ \bibnamefont {Marley}}, \bibinfo {author}
  {\bibfnamefont {T.}~\bibnamefont {Nakanishi}}, \bibinfo {author}
  {\bibfnamefont {R.~C.}\ \bibnamefont {Pardo}}, \bibinfo {author}
  {\bibfnamefont {K.~E.}\ \bibnamefont {Rehm}}, \bibinfo {author}
  {\bibfnamefont {D.}~\bibnamefont {Robertson}}, \bibinfo {author}
  {\bibfnamefont {R.}~\bibnamefont {Scott}}, \bibinfo {author} {\bibfnamefont
  {C.}~\bibnamefont {Schmitt}}, \bibinfo {author} {\bibfnamefont {X.~D.}\
  \bibnamefont {Tang}}, \bibinfo {author} {\bibfnamefont {R.}~\bibnamefont
  {Vondrasek}},\ and\ \bibinfo {author} {\bibfnamefont {A.}~\bibnamefont
  {Yokoyama}},\ }\href {https://doi.org/10.1126/science.1215510} {\bibfield
  {journal} {\bibinfo  {journal} {Science}\ }\textbf {\bibinfo {volume}
  {335}},\ \bibinfo {pages} {1614} (\bibinfo {year} {2012})},\ \Eprint
  {https://arxiv.org/abs/https://www.science.org/doi/pdf/10.1126/science.1215510}
  {https://www.science.org/doi/pdf/10.1126/science.1215510} \BibitemShut
  {NoStop}%
\bibitem [{\citenamefont {Gannoun}\ \emph {et~al.}(2011)\citenamefont
  {Gannoun}, \citenamefont {Boyet}, \citenamefont {Rizo},\ and\ \citenamefont
  {Goresy}}]{Gannoun2011}%
  \BibitemOpen
  \bibfield  {author} {\bibinfo {author} {\bibfnamefont {A.}~\bibnamefont
  {Gannoun}}, \bibinfo {author} {\bibfnamefont {M.}~\bibnamefont {Boyet}},
  \bibinfo {author} {\bibfnamefont {H.}~\bibnamefont {Rizo}},\ and\ \bibinfo
  {author} {\bibfnamefont {A.~E.}\ \bibnamefont {Goresy}},\ }\href
  {https://doi.org/10.1073/pnas.1017332108} {\bibfield  {journal} {\bibinfo
  {journal} {Proceedings of the National Academy of Sciences}\ }\textbf
  {\bibinfo {volume} {108}},\ \bibinfo {pages} {7693} (\bibinfo {year}
  {2011})},\ \Eprint
  {https://arxiv.org/abs/https://www.pnas.org/doi/pdf/10.1073/pnas.1017332108}
  {https://www.pnas.org/doi/pdf/10.1073/pnas.1017332108} \BibitemShut {NoStop}%
\bibitem [{\citenamefont {{Woosley}}\ and\ \citenamefont
  {{Howard}}(1978)}]{Woosley_APJS1978}%
  \BibitemOpen
  \bibfield  {author} {\bibinfo {author} {\bibfnamefont {S.~E.}\ \bibnamefont
  {{Woosley}}}\ and\ \bibinfo {author} {\bibfnamefont {W.~M.}\ \bibnamefont
  {{Howard}}},\ }\href {https://doi.org/10.1086/190501} {\bibfield  {journal}
  {\bibinfo  {journal} {Astroph. J. Suppl.}\ }\textbf {\bibinfo {volume}
  {36}},\ \bibinfo {pages} {285} (\bibinfo {year} {1978})}\BibitemShut
  {NoStop}%
\bibitem [{\citenamefont {{Audouze}}\ and\ \citenamefont
  {{Schramm}}(1972)}]{Audouze1972}%
  \BibitemOpen
  \bibfield  {author} {\bibinfo {author} {\bibfnamefont {J.}~\bibnamefont
  {{Audouze}}}\ and\ \bibinfo {author} {\bibfnamefont {D.~M.}\ \bibnamefont
  {{Schramm}}},\ }\href {https://doi.org/10.1038/237447a0} {\bibfield
  {journal} {\bibinfo  {journal} {Nature}\ }\textbf {\bibinfo {volume} {237}},\
  \bibinfo {pages} {445} (\bibinfo {year} {1972})}\BibitemShut {NoStop}%
\bibitem [{\citenamefont {{Prinzhofer}}\ \emph {et~al.}(1989)\citenamefont
  {{Prinzhofer}}, \citenamefont {{Papanastassiou}},\ and\ \citenamefont
  {{Wasserburg}}}]{Prinzhofer1989}%
  \BibitemOpen
  \bibfield  {author} {\bibinfo {author} {\bibfnamefont {A.}~\bibnamefont
  {{Prinzhofer}}}, \bibinfo {author} {\bibfnamefont {D.~A.}\ \bibnamefont
  {{Papanastassiou}}},\ and\ \bibinfo {author} {\bibfnamefont {G.~J.}\
  \bibnamefont {{Wasserburg}}},\ }\href {https://doi.org/10.1086/185536}
  {\bibfield  {journal} {\bibinfo  {journal} {Astrophys. J. Lett.}\ }\textbf
  {\bibinfo {volume} {344}},\ \bibinfo {pages} {L81} (\bibinfo {year}
  {1989})}\BibitemShut {NoStop}%
\bibitem [{\citenamefont {Prinzhofer}\ \emph {et~al.}(1992)\citenamefont
  {Prinzhofer}, \citenamefont {Papanastassiou},\ and\ \citenamefont
  {Wasserburg}}]{Prinzhofer1992}%
  \BibitemOpen
  \bibfield  {author} {\bibinfo {author} {\bibfnamefont {A.}~\bibnamefont
  {Prinzhofer}}, \bibinfo {author} {\bibfnamefont {D.}~\bibnamefont
  {Papanastassiou}},\ and\ \bibinfo {author} {\bibfnamefont {G.}~\bibnamefont
  {Wasserburg}},\ }\href
  {https://doi.org/https://doi.org/10.1016/0016-7037(92)90099-5} {\bibfield
  {journal} {\bibinfo  {journal} {Geochimica et Cosmochimica Acta}\ }\textbf
  {\bibinfo {volume} {56}},\ \bibinfo {pages} {797} (\bibinfo {year}
  {1992})}\BibitemShut {NoStop}%
\bibitem [{\citenamefont {Nyquist}\ \emph {et~al.}(1994)\citenamefont
  {Nyquist}, \citenamefont {Bansal}, \citenamefont {Wiesmann},\ and\
  \citenamefont {Shih}}]{Nyquist1994}%
  \BibitemOpen
  \bibfield  {author} {\bibinfo {author} {\bibfnamefont {L.~E.}\ \bibnamefont
  {Nyquist}}, \bibinfo {author} {\bibfnamefont {B.}~\bibnamefont {Bansal}},
  \bibinfo {author} {\bibfnamefont {H.}~\bibnamefont {Wiesmann}},\ and\
  \bibinfo {author} {\bibfnamefont {C.-Y.}\ \bibnamefont {Shih}},\ }\href
  {https://doi.org/https://doi.org/10.1111/j.1945-5100.1994.tb01102.x}
  {\bibfield  {journal} {\bibinfo  {journal} {Meteoritics}\ }\textbf {\bibinfo
  {volume} {29}},\ \bibinfo {pages} {872} (\bibinfo {year} {1994})},\ \Eprint
  {https://arxiv.org/abs/https://onlinelibrary.wiley.com/doi/pdf/10.1111/j.1945-5100.1994.tb01102.x}
  {https://onlinelibrary.wiley.com/doi/pdf/10.1111/j.1945-5100.1994.tb01102.x}
  \BibitemShut {NoStop}%
\bibitem [{\citenamefont {Rankenburg}\ \emph {et~al.}(2006)\citenamefont
  {Rankenburg}, \citenamefont {Brandon},\ and\ \citenamefont
  {Neal}}]{Rankenburg2006}%
  \BibitemOpen
  \bibfield  {author} {\bibinfo {author} {\bibfnamefont {K.}~\bibnamefont
  {Rankenburg}}, \bibinfo {author} {\bibfnamefont {A.~D.}\ \bibnamefont
  {Brandon}},\ and\ \bibinfo {author} {\bibfnamefont {C.~R.}\ \bibnamefont
  {Neal}},\ }\href {https://doi.org/10.1126/science.1126114} {\bibfield
  {journal} {\bibinfo  {journal} {Science}\ }\textbf {\bibinfo {volume}
  {312}},\ \bibinfo {pages} {1369} (\bibinfo {year} {2006})},\ \Eprint
  {https://arxiv.org/abs/https://www.science.org/doi/pdf/10.1126/science.1126114}
  {https://www.science.org/doi/pdf/10.1126/science.1126114} \BibitemShut
  {NoStop}%
\bibitem [{\citenamefont {Rauscher}\ and\ \citenamefont
  {Thielemann}()}]{NONSMOKER}%
  \BibitemOpen
  \bibfield  {author} {\bibinfo {author} {\bibfnamefont {T.}~\bibnamefont
  {Rauscher}}\ and\ \bibinfo {author} {\bibfnamefont {F.-K.}\ \bibnamefont
  {Thielemann}},\ }\href {http://www.nucastro.org/} {\bibinfo {title} {computer
  code {\sc{nonsmoker}}}}\BibitemShut {NoStop}%
\bibitem [{\citenamefont {Rauscher}\ \emph {et~al.}(2016)\citenamefont
  {Rauscher}, \citenamefont {Nishimura}, \citenamefont {Hirschi}, \citenamefont
  {Cescutti}, \citenamefont {Murphy},\ and\ \citenamefont
  {Heger}}]{Rauscher_MNRAS2016}%
  \BibitemOpen
  \bibfield  {author} {\bibinfo {author} {\bibfnamefont {T.}~\bibnamefont
  {Rauscher}}, \bibinfo {author} {\bibfnamefont {N.}~\bibnamefont {Nishimura}},
  \bibinfo {author} {\bibfnamefont {R.}~\bibnamefont {Hirschi}}, \bibinfo
  {author} {\bibfnamefont {G.}~\bibnamefont {Cescutti}}, \bibinfo {author}
  {\bibfnamefont {A.~S.~J.}\ \bibnamefont {Murphy}},\ and\ \bibinfo {author}
  {\bibfnamefont {A.}~\bibnamefont {Heger}},\ }\href
  {https://doi.org/10.1093/mnras/stw2266} {\bibfield  {journal} {\bibinfo
  {journal} {Mon. Not. R. Astron. Soc.}\ }\textbf {\bibinfo {volume} {463}},\
  \bibinfo {pages} {4153} (\bibinfo {year} {2016})}\BibitemShut {NoStop}%
\bibitem [{\citenamefont {Rauscher}(2010)}]{Rauscher_PRC2010_gamow}%
  \BibitemOpen
  \bibfield  {author} {\bibinfo {author} {\bibfnamefont {T.}~\bibnamefont
  {Rauscher}},\ }\href {https://doi.org/10.1103/PhysRevC.81.045807} {\bibfield
  {journal} {\bibinfo  {journal} {Phys. Rev. C}\ }\textbf {\bibinfo {volume}
  {81}},\ \bibinfo {pages} {045807} (\bibinfo {year} {2010})}\BibitemShut
  {NoStop}%
\bibitem [{\citenamefont {Psaltis}\ \emph {et~al.}(2022)\citenamefont
  {Psaltis}, \citenamefont {Arcones}, \citenamefont {Montes}, \citenamefont
  {Mohr}, \citenamefont {Hansen}, \citenamefont {Jacobi},\ and\ \citenamefont
  {Schatz}}]{Psaltis2022}%
  \BibitemOpen
  \bibfield  {author} {\bibinfo {author} {\bibfnamefont {A.}~\bibnamefont
  {Psaltis}}, \bibinfo {author} {\bibfnamefont {A.}~\bibnamefont {Arcones}},
  \bibinfo {author} {\bibfnamefont {F.}~\bibnamefont {Montes}}, \bibinfo
  {author} {\bibfnamefont {P.}~\bibnamefont {Mohr}}, \bibinfo {author}
  {\bibfnamefont {C.~J.}\ \bibnamefont {Hansen}}, \bibinfo {author}
  {\bibfnamefont {M.}~\bibnamefont {Jacobi}},\ and\ \bibinfo {author}
  {\bibfnamefont {H.}~\bibnamefont {Schatz}},\ }\href
  {https://doi.org/10.3847/1538-4357/ac7da7} {\bibfield  {journal} {\bibinfo
  {journal} {The Astrophysical Journal}\ }\textbf {\bibinfo {volume} {935}},\
  \bibinfo {pages} {27} (\bibinfo {year} {2022})}\BibitemShut {NoStop}%
\end{thebibliography}

\end{document}